\documentclass[prd,amsmath,amssymb,aps,floatfix,nofootinbib]{revtex4}
\pdfoutput=1 
\usepackage{graphicx,floatflt,amssymb,epsfig}
\usepackage[utf8]{inputenc}
\usepackage[inline]{enumitem}
\usepackage{mathrsfs}
\usepackage{amssymb}
\usepackage{amsmath}
\usepackage{color}
\usepackage{graphicx}
\usepackage{subfig}
\usepackage{multirow}
\usepackage{mathtools}
\usepackage{lscape}
\usepackage{float}
\usepackage{pstricks}
\usepackage[toc,page]{appendix}
\usepackage{pifont}
\usepackage{braket}
\usepackage{bbold}
\usepackage{ragged2e}
\usepackage[bookmarks=false,colorlinks]{hyperref}
\hypersetup{urlcolor=blue, citecolor=blue}
\usepackage{slashed}
\usepackage{rotating}

\usepackage{caption} 
\renewcommand{\arraystretch}{1.5}

{\newcommand{\lsim}{\mbox{\raisebox{-.6ex}{~$\stackrel{<}{\sim}$~}}}
	{\newcommand{\gsim}{\mbox{\raisebox{-.6ex}{~$\stackrel{>}{\sim}$~}}}

		\newcommand{\bmt}{\begin{pmatrix}}
			\newcommand{\emt}{\end{pmatrix}}
		\newcommand{\ba}{\begin{array}{c}}
			\newcommand{\ea}{\end{array}}
		\newcommand{\beq}{\begin{equation}}
		\newcommand{\eeq}{\end{equation}}
		\newcommand{\bea}{\begin{eqnarray}}
		\newcommand{\eea}{\end{eqnarray}}
		\newcommand{\nn}{\nonumber}
		\newcommand{\bi}{\begin{itemize}}
			\newcommand{\ei}{\end{itemize}}
		
		\newcommand{\baz}{\begin{array}{cc}}
			
			\newcommand{\mathsym}[1]{{}}

			\newcommand{\bt}{\begin{tabular}}
				\newcommand{\et}{\end{tabular}}

			\newcommand{\benu}{\begin{enumerate}}
				\newcommand{\eenu}{\end{enumerate}}
			
			\newcommand{\bav}{\begin{array}{cccc}}

\allowdisplaybreaks

\begin{document}
	
\title{\boldmath Constraining New Physics with Possible Dark Matter Signatures from a Global CKM Fit}

\author{Aritra Biswas,}
\email{iluvnpur@gmail.com}
\affiliation{Department of Physics, Indian Institute of Technology Guwahati, Assam 781039, India}
\author{Lopamudra Mukherjee,}
\email{mukherjeelopa@iitg.ac.in}
\affiliation{Department of Physics, Indian Institute of Technology Guwahati, Assam 781039, India}
\author{Soumitra Nandi,}
\email{soumitra.nandi@iitg.ac.in}
\affiliation{Department of Physics, Indian Institute of Technology Guwahati, Assam 781039, India}
\author{Sunando Kumar Patra}
\email{sunando.patra@gmail.com}
\affiliation{Department of Physics, Bangabasi Evening College, 19 Rajkumar Chakraborty Sarani, Kolkata, 700009, West Bengal, India}

\begin{abstract}
We constrain the parameters of a representative new physics model with possible dark matter (DM) signature from a global CKM fit analysis. The model has neutral quark current interactions mediated by a scalar, impacting the semileptonic and purely leptonic meson decays at one-loop. We take this opportunity to update the fit results for the Wolfenstein parameters and the CKM elements with and without a contribution from the new model using several other updated inputs. Alongside, we have analyzed and included in the CKM fit the $B\to D^*\ell\nu_{\ell}$ decay. The newly available inputs on the relevant form factors from lattice are included, and the possibility of new physics effects in $B\to D^*\ell\nu_{\ell}$ is considered. We obtain tight constraints on the relevant new physics parameters. We have studied the possible implications of this constraint on DM phenomenology. Apart from DM, the bounds are also applicable in other relevant phenomenological studies.
\end{abstract}

\maketitle
\section{Introduction}

The Standard Model of particle physics (SM) has emerged through theoretical and experimental discoveries, and has been tested extensively. Flavour physics has played an essential role in this development. Despite these successes, the SM fails to explain some key aspects of nature. For example, it can not provide a candidate for dark matter (DM), nor can it accommodate the observed baryon asymmetry. Therefore, extensions of the SM are formulated that address these issues by introducing new degrees of freedom beyond the SM. New particles or interactions introduced at a high scale could have a related shorter-distance interaction. The low energy observables will hence be useful in constraining the new physics (NP) parameter spaces. In the near future, they might play an essential role in the indirect detection of the new particles through deviations from the respective SM predictions.

The quark mixing matrix, also known as Cabibbo-Kobayashi-Maskawa (CKM) matrix, is important for understanding CP violation. The CKM matrix is a $3\times 3$ matrix, and a precise knowledge of the corresponding elements is essential. Following the Wolfenstein parametrization, four parameters are needed to define all the elements of the CKM matrix. Therefore, one of the important goals of the flavour studies is to constrain these four parameters using all the available measurements sensitive to the CKM matrix directly or indirectly.    

In the SM, the charged current interactions are the only flavour changing processes that occur at tree level, and the decay rates are directly sensitive to the square of the CKM elements. On the other hand, the FCNC processes are loop suppressed in the SM, and the corresponding amplitudes are sensitive to the product of CKM elements. Due to its simple and constrained structure in the SM, the weak processes are potentially sensitive to new interactions beyond the SM and hence can be a potent probe for models beyond the SM. It is necessary to measure the CKM parameters very precisely, and during the last few decades, extensive research has been performed at the BaBar, Belle and LHCb experiments. High-luminosity experiments like Belle-II have also become operational, and within a few years, we expect a wealth of precise data which will be useful to constrain NP model parameters. This paper will consider one such model that contributes to the semileptonic and purely leptonic decays at one-loop. Most of the inputs used to extract the Wolfenstein parameters and the related CKM elements are those coming from semileptonic and leptonic decays. At the moment, very precise measurements on the related observables are available which are hence beneficial in constraining the new model parameters contributing to these decays. We will analyze the constraints on the new parameters from observables related to the CKM measurements. Also, we will comment on the impact of such constraints on the DM phenomenology.

The simplest way to devise a dark matter model is by considering a scalar, fermionic or vector field obeying the SM gauge symmetries whose stability can be ensured by an additional discrete $\mathcal{Z}_2$ symmetry under which the DM is odd but all other SM particles are even. However, in order to annihilate into SM particles and give rise to the correct relic abundance, there has to be a mediator between the dark and the visible sectors. The interactions of the mediator with the visible sector may  include a non-zero vertex with the SM quark fields among others such that, the DM can scatter off a fixed target nuclei and be detected from any hint of nuclear recoil. However, such interactions might also impact important flavour physics observables which most of the dark matter analyses do not take into consideration. In this paper, we are going to investigate the constraints on the dark matter parameter space from flavour data in the context of a simple dark matter model.

\section{Simplified Model : Fermion Dark Matter with spin-0 Mediator}
It is common to use effective field theory (EFT) to describe the low energy effects of a high scale NP, in which non-renormalizable effective operators are added to the SM Lagrangian. However, due to the large energies accessible at the LHC, the interpretation of measurement using an effective theory approach may become questionable under certain circumstances. Therefore, the simplified model approaches \cite{Abdallah:2014hon,Abdallah:2015ter,DeSimone:2016fbz,Englert:2016joy,Albert:2016osu,
	Buchmueller:2017uqu}, which often contain both dark matter and mediator particles, have gained more relevance in the collider searches. In those models, the mediator provides the link between the visible SM particles and dark matter. By constructions, these simplified models do not contain all of the ingredients present in a ultra-violet (UV) complete model of dark matter. However, this approach helps characterize the dark matter production processes in UV-complete models without specifying the entire UV completion. There are different varieties of UV-complete models available in the literature, and it is not feasible to study all of them independently and constrain the parameter spaces. Also, the structures of such models are so rich that it is impossible to determine the underlying new dynamics unambiguously from a limited number of data. Hence, to constrain the NP parameters from the CKM measurements in this paper, we will consider a simplified model with a spin-0 mediator. A similar study for a simplified model with a vector mediator has been left for future work.        

For an illustration of our main objective, here we consider an extension of the SM by a singlet Dirac fermion dark matter $\chi$ and a spin-0 particle $S$. The DM decay is stabilized by imposing a discrete $\mathcal{Z}_2$ symmetry under which $\chi \to -\chi$ while all other particles remain even under the transformation. The most general renormalizable Lagrangian for such a model can be written as

\begin{equation}
\small
\mathcal{L} = \mathcal{L}_{SM} + \frac{1}{2} \bar{\chi}(i \slashed{\partial} - m_\chi) \chi - \frac{1}{2}(\partial_\mu S)^2 - \left[\bar{\chi}(g_s^\prime + i g_p^\prime \gamma_5) \chi + \bar{\psi}(g_s + i g_p \gamma_5)\psi \right] S - V(H,S) 
\label{eq:NPLag}
\end{equation}
where, $H$ denotes the SM Higgs doublet and $\psi$ denotes SM fermions. The scalar potential $V(H,S)$ can be of the form
\begin{equation}
\small
V(H,S) = \mu_H^2 H^\dagger H + \frac{1}{2}\lambda_H (H^\dagger H)^2 + \mu_1^3 S + \frac{\mu_S^2}{2} S^2 + \frac{\mu_3}{3!} S^3 + \frac{\lambda_S}{4!} S^4 + \lambda_1 (H^\dagger H) S + \frac{\lambda_2}{2}(H^\dagger H) S^2.
\label{eq:singlet-scalar-pot}
\end{equation}
For studies based on such models in the literature, see \cite{Schmeier:2013kda, Greljo:2013wja, Matsumoto:2018acr, Li:2016uph,Li:2018qip, Morgante:2018tiq}. Also, in our study, we mostly focus on effective DM interaction with SM quarks i.e $\psi \equiv q$. There are plenty of analyses on such leptophobic DM models in the perspective of LHC and indirect detection searches \cite{Li:2016uph,Li:2018qip, Morgante:2018tiq, Arina:2014yna,Abdallah:2015ter,Arina:2016cqj}.

Note that we do not attempt to construct a UV-complete model and constrain the interactions of the mediator with the DM and SM fermions only by the requirement of Lorentz invariance. In eq. \ref{eq:NPLag}, the mediator interaction with the SM fermions is renormalizable. However, it is not invariant under the SM gauge group. Hence, it is expected that the model will break down at a high energy scale $\Lambda$. One can assume that the origin of these interactions is some higher dimensional non-renormalizable operator suppressed by the power in $\Lambda$, which is invariant under the SM gauge group. The effective interaction mentioned above could also be generated from a UV complete model via the loop diagrams involving the heavy vector-like fermions that mix with the SM fermions, see for example ref.~\cite{Bauer:2016zfj}. There could be other ways as well; however, we have mentioned earlier that we are mostly interested in constraining the new couplings and not in the exact details of the origin of such interactions. Also, we should mention that for a UV complete description, one may need to add new states and interactions at the energy scale $\Lambda$. We will continue this discussion in the following section.


\begin{figure}[ht!]
\centering
\includegraphics[scale=0.7]{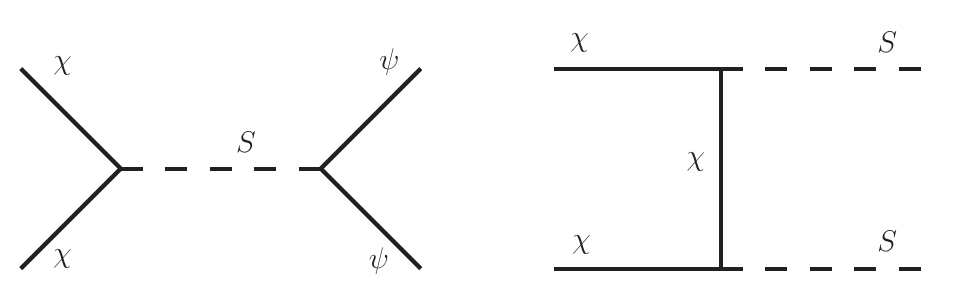}
\caption{Annihilation channels for the spin-0 mediated fermionic dark matter model under consideraion.}
\label{fig:DM-ann}
\end{figure}

In this analysis, we have considered only quarks and assumed universal coupling for all the quarks for the mediator interaction with the fermions. However, the above Lagrangian can produce large flavour violating effects due to the coupling of the mediator with SM quarks. Hence, invoking the prescription of minimal flavour violation (MFV), we scale the scalar and pseudoscalar couplings with the SM Yukawa couplings $y_{ij}$ as 
\beq
\mathcal{L}_{int}^{q} = \sum_{i,j} \bar{q}_i \frac{y_{ij}^q}{\sqrt{2}} (g_s + i g_p \gamma_5) q_j S
\label{eq:Lag1}
\eeq
where the sum runs over all quark flavours. In order to avoid FCNC, all flavour off-diagonal couplings are set to zero and the diagonal Yukawa couplings are given by $y_{ii}^q = \sqrt{2}m_f^q/v$ where $v = 246$ GeV, is the vacuum expectation value of the Higgs boson. Therefore, Eq,~\eqref{eq:Lag1} can be simplified as
\bea
\mathcal{L}_{int}^{q} &=& \bar{q}_i \frac{m_{q_i}}{v} (g_s+ i g_p \gamma_5) q_i S \\
&=& m_{q_i} \bar{q}_i (C_s+ i C_p \gamma_5) q_i S,
\label{eq:Lag2}
\eea
with $C_s = \frac{g_s}{v}$ and $C_p = \frac{g_p}{v}$, respectively.  Even though these couplings do not induce FCNC at tree level, one can have flavour-changing decays of K and B mesons induced by one-loop corrections leading to $s \to d S$ and $b \to d(s) S$ transitions. Constraints on the couplings in eq.~\eqref{eq:NPLag} from such decays have been studied in \cite{Dolan:2014ska} for mediators of mass $M_S \lsim 10$ GeV.

Following the Lagrangian given in Eqn.~\eqref{eq:NPLag}, it is evident that the dominant channel for DM annihilation will be the s-channel transition $\bar{\chi} \chi \to \bar{\psi} \psi$ shown by the Feynman diagram in the LHS of Fig.~\ref{fig:DM-ann}. There can also be a t-channel annihilation $\bar{\chi} \chi \to S S$ as shown in Fig.~\ref{fig:DM-ann} but for heavy scalars, that contribution will be much suppressed. The thermally averaged dark matter annihilation cross-section $\langle\sigma v\rangle$ is usually expressed as a partial-wave expansion in
powers of the square of the relative velocity between the annihilating particles as
\beq
\langle \sigma v \rangle = a + b\langle v^2 \rangle + d \langle v^4 \rangle + \cdots
\label{eq:sigmav}
\eeq
where $a,b,d$ are the leading s-wave, p-wave and d-wave contributions to the cross section respectively.
The dominant contribution to the s-channel DM annihilation rate for pure scalar interaction mediation is velocity suppressed due to the absence of s-wave terms. However, in  presence of the pseudoscalar coupling $g_p^{\prime}$, there is an enhancement in the annihilation cross-section due to the presence of an unsuppressed s-wave \cite{Arcadi:2017kky}. Also, there will be contributions to the direct detection cross section. The advantage of non-zero pseudoscalar interaction is that the WIMP-nucleon scattering cross-sections from such operators are spin dependent and velocity suppressed. This kind of pseudoscalar interactions helps to evade stringent bounds from present direct detection (DD) experimental searches.  The phenomenology of such pseudoscalar mediators have been extensively studied in \cite{Dolan:2014ska,No:2015xqa,Arcadi:2017wqi,Li:2018qip,Bell:2018zra}. While the pseudoscalar operators help weaken the direct detection scattering cross-section with a momentum suppression, they also amplify the chances of probing the WIMP at indirect detection experiments through initial/final state radiation or bremsstrahlung processes \cite{Flores:1989ru, Bell:2011if, Bell:2011eu, Bell:2017irk, Kumar:2016mrq, Clark:2019nby}. On the other hand, the only way to obtain a spin-independent direct detection cross-section is to have a non-zero scalar-scalar effective interaction i.e $C_s, g_s^\prime \neq 0$.

\section{Contributions in $d_i \to u_j \ell\nu_{\ell}$ decays}

\begin{figure}[t]
	\centering
	\includegraphics[scale=0.7]{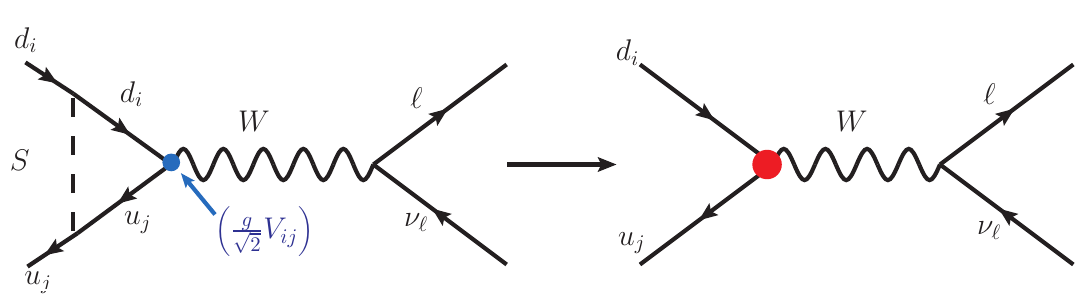}
	\caption{Loop correction to the $d_i \to u_j  W$ vertex in the presence of a real scalar $S$. The vertex modification will have direct impact on the vertex CKM factor $V_{ij}$.}
	\label{fig:vertex-corr}
\end{figure}

In the SM, the $d_i \to u_j \ell\nu_{\ell}$ transitions are tree level processes mediated by $W$-boson. Therefore, the $d_i \to u_j  W$ vertex has a $V-A$ structure i.e $\gamma^\mu (1-\gamma_5)$. In the previous section, in Eq.~\ref{eq:Lag2} we define a Lagrangian which contains interactions of SM fermions with the scalar $S$. Note that this type of interaction will affect the SM charged current vertex $\bar d_j \gamma^{\mu}(1-\gamma_5) u_i W_{\mu}$ at one loop, resulting in new contributions in the semileptonic or purely leptonic decay rates $\Gamma_{(d_j \to u_i\ell\nu_{\ell})}$ $(\ell =leptons)$. The representative diagram is shown in Fig.~\ref{fig:vertex-corr}, wherein these decays receive vertex corrections from the heavy scalar exchanges in the loop. The CKM element $V_{ij}$ appears as a vertex factor of the charged current interactions in the SM. As will be shown in the next subsection, the corrections due to NP have a direct impact on the vertex factors, which in this case are the CKM elements multiplied by the $SU(2)_L$ gauge coupling : $V_{ij}\frac{g}{\sqrt{2}}$. The vertex correction shown in Fig.~\ref{fig:vertex-corr} may introduce additional operators other than $(V-A)$ type.

The most general effective Hamiltonian for the $d_i \to u_j \ell \nu$ processes can be expressed as \cite{HFLAV2019,PDG2020} : 
\beq
\mathcal{H}_{\text{eff}}^{d_i \to u_j} = \frac{4 G_F}{\sqrt{2}} V_{ij} \left[(\delta_{\ell \ell} + C_{V_1}^\ell) \mathcal{O}_{V_1}^\ell + C_{V_2}^\ell \mathcal{O}_{V_2}^\ell +C_{S_1}^\ell \mathcal{O}_{S_1}^\ell + C_{S_2}^\ell \mathcal{O}_{S_2}^\ell + C_{T}^\ell \mathcal{O}_{T}^\ell \right]
\label{eq:Heff}
\eeq
where $C_X^\ell$ ($X= V_1, V_2, S_1, S_2, T$) are the Wilson coefficients (WCs) corresponding to the operator basis
\bea
\mathcal{O}_{V_1}^{\ell} &=& (\bar{u}_{jL} \gamma^\mu d_{iL})(\bar{\ell}_L \gamma_\mu \nu_{L}), \nonumber \\ 
\mathcal{O}_{V_2}^\ell &=& (\bar{u}_{jR} \gamma^\mu d_{iR})(\bar{\ell}_L \gamma_\mu \nu_{L}), \nonumber \\ 
\mathcal{O}_{S_1}^\ell &=& (\bar{u}_{jL} d_{iR})(\bar{\ell}_R \nu_{L}), \\ 
\mathcal{O}_{S_2}^\ell &=& (\bar{u}_{jR} d_{iL})(\bar{\ell}_R \nu_{ L}), \nonumber \\ 
\mathcal{O}_{T}^\ell &=& (\bar{u}_{jR} \sigma^{\mu \nu} d_{iL})(\bar{\ell}_R \sigma_{\mu \nu} \nu_{ L}). \nonumber
\label{eq:effops}
\eea
There are no lepton flavour violating vertices in the Lagrangian under consideration Eq.~\ref{eq:Lag2}. Hence, for all practical purposes, we can remove the suffix $\ell$ in the operator basis and write $C_X^\ell \equiv C_X$. Note that in the SM, only $\mathcal{O}_{V_1}$ contributes at the tree level. Along with $\mathcal{O}_{V_1}$, the rest of the operators may appear by themselves or as combinations in different NP scenarios. The WC $C_X$ incorporates the NP effects in these decays, and therefore in the SM, $C_X =0$. 

The detailed mathematical expressions of the decay rate distributions for the exclusive semileptonic $P \to M^{(*)} \ell\nu_{\ell}$ and purely leptonic $P\to \ell\nu_{\ell}$ decays can be seen from ref. \cite{HFLAV2019} where $P$ and $M$ are the pseudoscalar mesons, and $M^*$ is a vector meson. The semileptonic and purely leptonic decays rates are directly proportional to the vertex factors.  Here, we would like to mention that most of the CKM elements, like $|V_{ud}|$, $|V_{cd}|$, $|V_{us}|$, $|V_{cs}|$, $|V_{ub}|$, $|V_{cb}|$, are extracted from the semileptonic and purely leptonic (few cases) $d_i \to u_j \ell\nu_{\ell}$ decays with $\ell = \mu $, or $e$. The underlying assumption is that these decays with the light leptons will be less sensitive to any NP effect. 
The measured decay rates, along with some other inputs from lattice (decay constants and form factors), are useful probes for the CKM elements $|V_{ij}|$. In the presence of new four-fermi operators in accordance to Eqn.~\eqref{eq:Heff}, the decay rates will be modified. If only the vertex factor is modified, then the extracted values of the $|V_{ij}|$ can be directly used to constrain the new couplings, else, we need to fit the decay rates themselves. In the following subsections, we will discuss this in detail. 

Also, it is important to mention that all these CKM elements are extracted with reasonably good precision. For example $|V_{ud}|$ and $|V_{cs}|$ are known with an error $\approx 0.01\%$ while $|V_{us}|$ and $|V_{cd}|$ are known with an accuracy of $0.1\%$. The $|V_{ub}|$ and $|V_{cb}|$ are relatively less precisely known. Therefore it is natural to expect tight constraints on the new couplings $C_s$ and $C_p$ from an analysis of the CKM observables, purely leptonic and exclusive semileptonic decay rates respectively. Note that $|V_{ub}|$ and $|V_{cb}|$ are also extracted from semileptonic inclusive decays. We do not consider the inputs from inclusive decays to constrain the new couplings. The extraction of $|V_{cb}|$ from the inclusive decays requires a complex fit to the respective decay rate and moments. Considering leading order power correction up to order $1/m_b^5$ one needs to fit 18 non-perturbative matrix elements alongside $|V_{cb}|$, $m_b$ and $m_c$ \cite{Gambino:2013rza,Alberti:2014yda,Gambino:2016jkc}. Due to the insufficient number of inputs, model assumptions are used in the fit for a couple of higher-order non-perturbative matrix elements. On top of this, unknown higher-order corrections to the non-perturbative matrix elements are also relevant to improve the precision of $|V_{cb}|$, which may be small compared to the known ones but missing at the moment. The study of NP contributions in the inclusive decays is a dedicated project where we need to calculate analytically the NP contributions in the decay rates, as well as in the hadronic, leptonic and the $q^2$ moments and then one needs to do a simultaneous fit.  Here, the dependencies of these observables to the new physics parameters will not be simple. Hence extractions of new physics information won't be very clean. The situation is even worse in the inclusive determination of the $|V_{ub}|$. Here, the results are completely dependent on the QCD modelling of the non-perturbative shape functions on top of the non-perturbative matrix elements. There are four different methods used in the literature to model the shape function, and the extracted values of $|V_{ub}|$ in each of these methods do not exactly agree with each other \cite{HFLAV2020,PDG2020}. Therefore, it is natural to expect that before constraining NP from inclusive $b\to u\ell\nu_{\ell}$, we need to understand the underline methodology first.

On the contrary, we have sufficient number of inputs from experiments and lattice for the exclusive determinations $|V_{ub}|$ and $|V_{cb}|$. Note that the respective rates have a very simple dependence on the new Wilson coefficients in these decays. Hence, the extractions of these coefficients will be relatively clean compared to that from the inclusive decays, given the complexity of the fit in the inclusive decays, as discussed above. Numerically, we do not expect any changes in the allowed parameter spaces of new coefficients from the inclusion of inclusive decays. Also, it should be noted that the majority of the other inputs used in CKM fit analysis have relatively better precision than $|V_{ub}|$ and $|V_{cb}|$ from inclusive decays.

\subsection{Effective vertex}
As mentioned earlier, in the SM, the coupling strength for the $d_i\to u_j W$ charged current interation is given by $\frac{i g V_{ij}}{\sqrt{2}}$ and the interaction is of the type $(V-A)$. However, the one-loop correction of this charged current vertex due to the interaction given in Eq.~\ref{eq:Lag2} introduces one new $(V+A)$ type interaction in addition to the original $(V-A)$ type interaction. The corresponding Feynman diagram can be seen from Fig.~\ref{fig:vertex-corr}, and the effective charged current interaction can be written as:
\begin{align}
{\cal L}^{eff}_{d_i\to u_j W} &= \frac{i g V_{ij}}{2\sqrt{2}}\left[ C_L {\bar u_j}\gamma_{\mu}(1-\gamma_5) d_i  + C_R {\bar u_j} \gamma_{\mu}(1 +\gamma_5)d_i \right] W^{\mu} \nn\\
 & = \frac{i g V_{ij}}{2\sqrt{2}}\left[ C_L \mathcal{O}_L + C_R \mathcal{O}_R \right] W^{\mu}.
 \label{eq:effvertex}
\end{align}  
Here, the effects of NP coming from the loop corrections are introduced in the coefficients $C_L$ and $C_R$, respectively. Hence, we can say that at the tree level (pure SM) $C_L =1$ and $C_R=0$. We have performed the calculation in a unitary gauge using dimensional regularization and find that the one loop contribution to charged current vertex is in general divergent.  
\begin{figure}[t]
	\centering
	\includegraphics[scale=0.7]{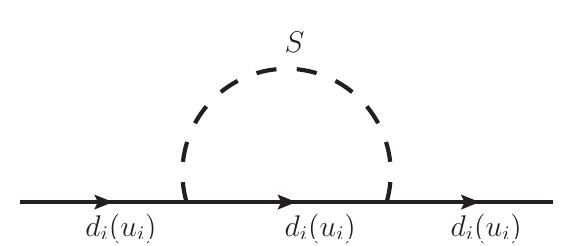}
	\caption{Quark self energy corrections in the presence of the new interaction given in Eq.~\ref{eq:Lag2}.}
	\label{fig:self-corr}
\end{figure}
The loop factor $C_L$ does not receive any $\frac{1}{\epsilon}$ pole from diagram \ref{fig:vertex-corr}, however, $C_R$ has a pole. 
Note that there will be a contribution to $C_L$ from the vertex counter term. The relevant part of the counter term can be obtained from the wave function renormalization which we have calculated from the quark self-energy correction diagram given in Fig. \ref{fig:self-corr}. 
For the renormalization we have followed the $\overline{MS}$-scheme. 

Note that the overall contributions in $C_L$ and $C_R$ will be divergent, and we don't have any additional contributions to remove these divergences. It is well-known that in simplified models as defined above the
one-loop contributions to flavour-changing transitions are in general UV divergent, see for example the following refs.~\cite{Batell:2009jf,Freytsis:2009ct,Arcadi:2017kky}. To make the theory renormalizable, one may need to add new states with tree-level charged current interactions with the SM fermions. However, the available data will constrain such an interaction in general. As mentioned earlier, such divergences reflect the dependence of our results on the suppression scale $\Lambda$ (the scale at which new states might appear). Therefore, we will interpret our model as an effective theory below some new physics scale $\Lambda$ with the following replacement: $1/\epsilon + Log(\mu^2/m^2) \to Log(\Lambda^2/m^2)$. In a UV-complete theory, the additional new physics at the scale $\Lambda$ is expected to cancel the divergences present. Here, we take an optimistic view and assume that the new high-scale (higher-dimensional operators) contributions in the low-energy observables will have a negligible impact on our analysis. The renormalization group evolution (RGE) over the energy range that we have considered here will not change the coupling structure significantly.  

After obtaining the contributions in $C_L$ and $C_R$ and integrating out the $W$ field from the diagram of Fig. \ref{fig:vertex-corr}, we obtain the following effective Hamiltonian
\beq
\mathcal{H}_{\text{eff}}^{d_i \to u_j} = \frac{4 G_F}{\sqrt{2}} V_{ij} \left[(1 + C_{V_1}) \mathcal{O}_{V_1} + C_{V_2} \mathcal{O}_{V_2}  \right],
\label{eq:Hefff}
\eeq
where the operators are defined in Eq.~\eqref{eq:effops}. The WCs $C_{V_1}$ and $C_{V_2}$ will be obtained from $C_L$ and $C_R$. In the leading-log approximation, the WCs can be expressed as,
\beq
C_{V_1}^{q_i\to q_j} = -\frac{C_T}{64\pi^2}m_{q_i}^2 \log \frac{\Lambda^2}{m_{q_i}^2}
\label{eq:CV1}
\eeq
\beq
C_{V_2}^{q_i\to q_j} = -\frac{C_T}{32\pi^2}m_{q_i} m_{q_j} \log \frac{\Lambda^2}{m_{q_i}^2}
\label{eq:CV2}
\eeq
where $C_T \equiv (C_s^2 + C_p^2) = (g_s^2 + g_p^2)/v^2$. We have dropped the other finite contributions, which are small effects as compared to the one given in the above equations. 
It is important to note that the loop contribution is zero in the massless quark limit for light quarks, such as $u,d,s$, etc. In the above we have chosen the scale $\Lambda$ as the UV cutoff while the IR scales will be decided by the respective decaying quark masses.

We want to point out that, in the SM we can have similar vertex corrections with the scalar $S$ replaced by the SM Higgs or by a $Z$ boson in Fig.~\ref{fig:vertex-corr}. We can parametrize such correction as $\delta C_{V_1}^{SM}$ which represent a small shift from $C_{V_1}^{SM} = 1$. For SM Higgs, there won't be any contribution in $C_{V_2}$ and the contribution in $\delta C_{V_1}^{SM}$ is $\lsim 10^{-8}$. For SM Z-boson, the contribution to both $\delta C_{V_1}^{SM}$ and $C_{V_2}$ are negligibly small as compared to the new contribution in $C_{V_2}$. We hence drop any such contribution in our analysis since they have a negligible impact on our findings.   

\subsection{Contributions in the decays: semileptonic and leptonic}

Using the effective Hamiltonian given in Eqn.~\eqref{eq:Hefff}, the differential decay rate for the $P \to M \ell \nu_\ell$ transition is written as~\cite{Sakaki:2013bfa}
\beq
\small
\frac{d\Gamma (P \to M \ell \nu_\ell)}{dq^2} = \frac{G_F^2 |V_{ij}|^2}{\pi^3 m_P^3} q^2 \sqrt{\lambda_M (q^2)} \left(1-\frac{m_\ell^2}{q^2}\right) |1+C_{V_1} + C_{V_2}|^2 \left \{ \left(1+\frac{m_\ell^2}{2q^2}\right) {H_{V,0}^{s}}^2 + \frac{3}{2}\frac{m_\ell^2}{q^2} {H_{V,t}^s}^2 \right \},
\eeq
while that for $P \to M^* \ell \nu_\ell$ is
\beq
\small
\begin{aligned}
\frac{d\Gamma (P \to M^* \ell \nu_\ell)}{dq^2} &= \frac{G_F^2 |V_{ij}|^2}{\pi^3 m_P^3} q^2 \sqrt{\lambda_{M^*} (q^2)} \left(1-\frac{m_\ell^2}{q^2}\right) \times \bigg \{  \\&
(|1+C_{V_1}|^2 + |C_{V_2}|^2) \left[ \left(1+\frac{m_\ell^2}{2q^2}\right)( H_{V,+}^2 + H_{V,-}^2 +H_{V,0}^2) + \frac{3}{2}\frac{m_\ell^2}{q^2} H_{V,t}^2 \right] \\& 
-2 \mathcal{R}e[(1+C_{V_1})C_{V_2}^{ *}]\left[ \left(1+\frac{m_\ell^2}{2q^2}\right)( H_{V,0}^2 + 2H_{V,+}~H_{V,-}) + \frac{3}{2}\frac{m_\ell^2}{q^2} H_{V,t}^2 \right] \bigg \}.
\end{aligned}
\eeq
The helicity amplitudes are written in terms of the QCD form factors as given below 
\begin{subequations}
	\begin{align}
	H_{V,0}^s(q^2) &  = \sqrt{\lambda_M(q^2) \over q^2} f_+(q^2) \,, \\
	& \nonumber \\
	H_{V,t}^s(q^2) &  = {m_P^2-m_M^2 \over \sqrt{q^2}} f_0(q^2).
	\end{align}
\end{subequations}
and
\begin{subequations}
	\begin{align}
	H_{V,\pm}(q^2)  = &  (m_P+m_{M^*}) A_1(q^2) \mp { \sqrt{\lambda_{M^*}(q^2)} \over m_P+m_{M^*} } V(q^2) \,, \\
	& \nonumber \\
	H_{V,0}(q^2) & = { m_P+m_{M^*} \over 2m_{M^*} \sqrt{q^2} } \left[-(m_P^2-{m_{M^*}}^2-q^2) A_1(q^2) \right.\left. + { \lambda_{M^*}(q^2) \over (m_P+m_{M^*})^2 } A_2(q^2) \right] \,, \\
	& \nonumber \\
	H_{V,t}(q^2) &  = -\sqrt{ \lambda_{M^*}(q^2) \over q^2 } A_0(q^2) \, .
	\end{align}
\end{subequations} 
The branching fraction for $P\to\ell\nu_{\ell}$ corresponding to the same Hamiltonian is:
\begin{equation}\label{eq:brlep}
\begin{split}
\mathcal{B}(P\to\ell\nu_{\ell}) = & \frac{\tau_P}{8\pi}m_P m_\ell^2 f_P^2 G_F^2 \left(1-\frac{m_\ell^2}{m_P^2}\right)^2 \left|V_{ij}(1+C_{V_1}-C_{V_2}) \right|^2.
\end{split}
\end{equation}

From the above decay rate distributions, we can see that the new contributions to $P \to M\ell\nu_{\ell}$ and $P\to \ell\nu_{\ell}$ decays will modify only the vertex from $|V_{ij}| \rightarrow |V_{ij}^\prime| = |V_{ij} (1+ C_{V_1} \pm C_{V_2})|$, respectively. However, in $P \to M^*\ell\nu_{\ell}$ transitions the new contributions will modify the $q^2$ distribution. Therefore, the CKM elements $|V_{ij}^{\prime}|$ extracted from purely leptonic or $P\to M\ell\nu_{\ell}$ decays, can be directly used to constrain the new parameters along with the Wolfenstein parameters: $A, \lambda, \rho$ and $\eta$ with which we need to parametrize $|V_{ij}|$. Note that $|V_{cb}|$ is extracted from both $B \to D \ell \nu_\ell$ and $B \to D^* \ell \nu_\ell $ decays. Hence, to extract the Wolfenstein parameters along with the new parameters from $B \to D^* \ell \nu_\ell $ decays, we need to redo the fit to the experimental data. We will discuss the relevant details in the next section.

\section{Numerical Analysis and Results}

\subsection{$B \to D^{*}\ell \nu$ Observables}\label{subsec:BtoDst}
As pointed out in the previous section, for the NP scenario under consideration we need to fit the decay rate distributions of $B\to D^*\ell\nu_{\ell}$ decays to extract the CKM parameters along with the NP parameters. The methodology of this fit will be similar to the one given in refs.~\cite{Jaiswal:2017rve,Jaiswal:2020wer} with very recent updates from the Fermilab Lattice Collaboration~\cite{FermilabLattice:2021cdg}. For the first time, they have provided the $B \to D^*$ form factors at non-zero recoils. They provide a set of synthetic data based on the Boyd-Grinstein-Lebed (BGL) parametrization~\cite{Boyd:1997kz} of the form factors truncated at $N =2$ at three $w$ values, $\{1.03,1.10,1.17\}$, along with their correlations. We have used these data points in our analysis. In accordance to our previous work, we have utilized the untagged dataset for the four-fold decay distribution corresponding to $B \to D^{*}\ell \nu$ by the Belle collaboration~\cite{Belle:2018ezy}. We have also used the $B \to D^*$ form factors at $q^2 = 0$ from QCD Light-Cone Sum Rules (LCSR)~\cite{Gubernari:2018wyi}. Additionaly, the Fermilab/MILC lattice input, $h_{A_1}(1) = 0.906(13)$~\cite{FermilabLattice:2014ysv}, allows us to efficiently constrain the form factor parameter $a_0^f$ and hence, $|V_{cb}|$. In our analysis we have not considered the dataset of unfolded differential decay rates of four kinematic variables for $\overline{B}^0 \to D^{*+} \ell^- \bar{\nu}_\ell$ provided by Belle in 2017 \cite{Abdesselam:2017kjf} since the data remains unpublished till date. However, in Appendix.~\ref{sec:Belle2017}, we have provided our fit results by including this dataset. As we will see later, the inclusion of this data does not affect our fit results much.

\begin{table}[t]
	\centering
	\renewcommand{\arraystretch}{1.4}
	\begin{tabular}{|c|c|c|c|c|}
		\hline
		\multirow{2}{*}{Dataset} & \multicolumn{2}{|c|}{Fit Quality} & \multirow{2}{*}{Parameter} & \multirow{2}{*}{Fit Result} \\
		\cline{2-3} 
		& $\chi^2/dof$ & p-Value & &\\
		\hline
		& \multirow{11}{*}{$52.82/45$} & \multirow{11}{*}{$19.75\%$} & $|V_{cb}|$ & $38.69 (79) \times 10^{-3} $ \\
		& & & $a_{0}^f$ & $0.0123 (1)$ \\
		& & & $a_{1}^f$ & $0.0222 (96)$ \\
		& & & $a_{2}^f$ & $-0.522 (196)$ \\
		\multirow{2}{*}{Belle'18 \cite{Belle:2018ezy} + $h_{A_1} (1)$ \cite{FermilabLattice:2014ysv}} & & & $a_{0}^g$ & $0.0318 (10)$ \\
		+ LCSR \cite{Gubernari:2018wyi} + Lattice \cite{FermilabLattice:2021cdg} & & & $a_{1}^g$ & $-0.133 (63)$ \\
		& & & $a_{2}^g$ & $-0.62 (146)$ \\
		& & & $a_{1}^{\mathcal{F}_1}$ & $0.0021 (15)$ \\
		& & & $a_{0}^{\mathcal{F}_2}$ & $0.0515 (12)$ \\
		& & & $a_{1}^{\mathcal{F}_2}$ & $-0.149 (59)$ \\
		& & & $a_{2}^{\mathcal{F}_2}$ & $0.987 (932)$ \\
		\hline
	\end{tabular}
	\caption{Fit result for the frequentist analysis of the mentioned $B \to D^* \ell \bar{\nu}_\ell$ dataset for the SM scenario.}
	\label{tab:BDstlnu-SM}
\end{table}
The four form factors relevant for $B \to D^* \ell \nu_\ell$ decay are $\mathcal{F}_i = \{f(z)$, $g(z)$, $\mathcal{F}_1 (z), \mathcal{F}_2 (z)\} $. In the BGL method of parametrization, these form factors can be expressed as a series expansion in $z$ as
\beq
\mathcal{F}_i (z) = \frac{1}{P_i (z) \phi_i (z)} \sum_{j=0}^{N} a_{j}^{\mathcal{F}_i} z^j,
\label{eq:FF-BGL}
\eeq
where $z$ is related to the recoil angle $w$ as
\beq
z = \frac{\sqrt{w+1}-\sqrt{2}}{\sqrt{w+1}+\sqrt{2}}.
\eeq
The recoil angle is related to the momentum transfer $q^2$ as $q^2 = m_B^2 + m_{D^*}^2 - 2 m_B m_{D^*} w$.
The functions $P_i (z)$, called the Blaschke factors,  are given by
\beq
P_i(z) = \prod_p \frac{z-z_p}{1 - z z_p},
\label{eq:Blaschke-fact}
\eeq
which are used to eliminate the poles at $z=z_p$ where,
\beq
z_p = \frac{\sqrt{(m_B + m_{D^*})^2 - m_P^2} - \sqrt{4 m_B m_{D^*}}}{\sqrt{(m_B + m_{D^*})^2 - m_P^2} + \sqrt{4 m_B m_{D^*}}}.
\label{eq:zp}
\eeq
Here $m_P$ denotes the pole masses and can be looked up from \cite{Bigi:2017jbd}. The outer functions $\phi_i (z)$ are chosen to be
\bea
\phi_f &=& \frac{4r}{m_B^2} \sqrt{\frac{n_I}{6\pi \chi_{1^+}^T (0)}} \frac{(1+z)(1-z)^{3/2}}{\left[(1+r)(1-z) + 2\sqrt{r}(1+z)\right]^4}, \nonumber \\
\phi_g &=& 16r^2 \sqrt{\frac{n_I}{3\pi \tilde{\chi}_{1^-}^T (0)}} \frac{(1+z)^2(1-z)^{-1/2}}{\left[(1+r)(1-z) + 2\sqrt{r}(1+z)\right]^4},\\
\phi_{\mathcal{F}_1} &=& \frac{4r}{m_B^3} \sqrt{\frac{n_I}{6\pi \chi_{1^+}^T (0)}} \frac{(1+z)(1-z)^{5/2}}{\left[(1+r)(1-z) + 2\sqrt{r}(1+z)\right]^5}, \nonumber \\
\phi_{\mathcal{F}_2} &=& 8\sqrt{2}r^2 \sqrt{\frac{n_I}{\pi \tilde{\chi}_{1^+}^L (0)}} \frac{(1+z)^2 (1-z)^{-1/2}}{\left[(1+r)(1-z) + 2\sqrt{r}(1+z)\right]^4} \nonumber
\label{eq:outer-func}
\eea
where $r = m_{D^*}/m_B$ and the other inputs can be found in \cite{Bigi:2017jbd}. Therefore, for $N = 2$, there are twelve coefficients, $a_{j}^{\mathcal{F}_i} $ for the four form factors. These coefficients satisfy the following weak unitarity constraints :
\beq
\sum_{j=0}^{N} (a_j^{g})^2 < 1,~~\sum_{j=0}^{N} (a_j^{f})^2 + (a_j^{\mathcal{F}_1})^2  < 1, ~~\sum_{j=0}^{N} (a_j^{\mathcal{F}_2})^2  < 1.
\label{eq:FF-unity-constr}
\eeq
Furthermore, there are two kinematical constraints on the form factors, one each at zero and maximum recoil :
\bea
\mathcal{F}_1 (1) &=& m_B (1-r) f(1),\\
\mathcal{F}_2 (w_{max}) &=& \frac{1+r}{m_B^2 (1+w_{max}) (1-r)r} \mathcal{F}_1 (w_{max}).
\label{eq:FF-kin-constr}
\eea
We consider these constraints in our analysis to remove two of the BGL coefficients from the theory. In the limit of massless leptons, the decay distribution becomes insensitive to the form factor $\mathcal{F}_2$. Hence, only 8 independent form factor coefficients are required to fit the theory to the data. For the numerical analysis presented here, we perform a maximum likelihood estimation of the parameters using Optex, a Mathematica based package. The fit results are provided in Table.~\ref{tab:BDstlnu-SM}. The value of $|V_{cb}|$ is extremely consistent with the one obtained in \cite{FermilabLattice:2021cdg}\footnote{ In Table.~\ref{tab:BDstlnu-SM-2} in Appendix.~\ref{sec:Belle2017}, we check the fit by additionally including the 2017 Belle data. We find that the value of $|V_{cb}|$ is consistent with the value obtained without this dataset at $1\sigma$ CL.}. In the following section we will utilize this value of $|V_{cb}|$ for a global CKM fit without NP.  

\begin{table}[t]
	\centering
	\renewcommand{\arraystretch}{1.9}
	\resizebox{\textwidth}{!}{
		\begin{tabular}{c|c|c|c|c|c|c}
			\cline{2-7}
			& \multirow{2}{*}{$A$} & \multirow{2}{*}{$\lambda$} & \multirow{2}{*}{$\bar{\rho}$} & \multirow{2}{*}{$\bar{\eta}$} & \multicolumn{2}{c|}{Fit Quality} \\
			\cline{6-7}
			&&&&& $\chi^2/dof$ & \multicolumn{1}{c|}{p-Value}\\
			\hline
			\multicolumn{1}{|c|}{\textit{CKMFitter'19}} & $0.8235^{+0.0056}_{-0.0145}$ & $0.224837^{+0.000251}_{-0.000060}$ & $0.1569^{+0.0102}_{-0.0061}$ & $0.3499^{+0.0079}_{-0.0065}$ & - & \multicolumn{1}{c|}{-}\\
			\hline
			\multicolumn{1}{|c|}{\textit{Our Result}} & $0.8205 \pm 0.0075$ & $0.22462 \pm 0.00031$ & $0.1607 \pm 0.0093$ & \multicolumn{1}{c|}{$0.3558 \pm 0.0088$} & $34.18/23$ & \multicolumn{1}{c|}{$6.26\%$}\\
			\hline
			\multicolumn{1}{|c|}{\textit{Updated 2021 Results}} & $0.8178 \pm 0.0070$ & $0.22498 \pm 0.00029$ & $0.1734 \pm 0.0092$ & \multicolumn{1}{c|}{$0.374 \pm 0.011$} & $37.25/25$ & \multicolumn{1}{c|}{$8.37\%$} \\
			\hline
		\end{tabular}}
		\caption{Comparison of the best fit estimates of the Wolfenstein parameters by the CKMFitter group and our group from the global CKM fit in the SM framework. The two results are consistent with each other within $1\sigma$ limit of the errors. We also provide the $\chi^2/dof$ and the goodness of fit for our fit results. The last row contains the best parameter estimates of the global scenario with the most updated inputs.}
		\label{tab:CKM-fit}
	\end{table}

\subsection{CKM Fit}
\begin{table}
	\centering
	\renewcommand{\arraystretch}{1.2}
	\begin{tabular}{|c|c|c|}
		\hline
		Observable & Value & Reference \\
		\hline
		$|V_{ud}|$ (nucl) & $0.97420 \pm 0.00021$ & \cite{Hardy:2018zsb}\\
		$|V_{us}|f_+^{K \to \pi}(0)$ & $0.2165 \pm 0.0004$ & \cite{Moulson:20174Z}\\
		$|V_{cd}|_{\nu N}$ & $0.30 \pm 0.011$ & \cite{PDG2020}\\
		$|V_{cs}|_{W \to c\bar{s}}$ & $0.94^{+0.32}_{-0.26} \pm 0.13$ & \cite{PDG2020}\\
		$|V_{ub}|_{excl}$ & $(3.91 \pm 0.13)\times 10^{-3}$ & \cite{Biswas:2021qyq,Biswas:2021cyd}\\
		$|V_{ub}|_{incl}$ & $(4.10^{+0.09}_{-0.22} \pm 0.15)\times 10^{-3}$ & \cite{Capdevila:2021vkf}\\
		$|V_{cb}|_{B\to D}$ & $(40.84 \pm 1.15) \times 10^{-3}$ & \cite{Jaiswal:2017rve} \\
		$|V_{cb}|_{B\to D^*}$ & $(38.69 \pm 0.79) \times 10^{-3}$ & this work \\
		$|V_{cb}|_{incl}$ & $(42.16 \pm 0.50) \times 10^{-3}$ & \cite{Bordone:2021oof}\\
		$\mathcal{B}(\Lambda_p \to p \mu^- \bar{\nu}_\mu)_{q^2 > 15}/\mathcal{B}(\Lambda_p \to \Lambda_c \mu^- \bar{\nu}_\mu)_{q^2 > 7}$ & $(0.947 \pm 0.081)\times 10^{-2}$ & \cite{LHCb:2015eia} \\
		$\mathcal{B}(B^- \to \tau^- \bar{\nu}_\tau)$ & $(1.09 \pm 0.24) \times 10^{-4}$ & \cite{HFLAV2019} \\
		$\mathcal{B}(D_s^- \to \mu^- \bar{\nu}_\mu)$ & $(5.51 \pm 0.16) \times 10^{-3}$ & \cite{HFLAV2019}\\
		$\mathcal{B}(D_s^- \to \tau^- \bar{\nu}_\tau)$ & $(5.52 \pm 0.24) \times 10^{-2}$ & \cite{HFLAV2019} \\
		$\mathcal{B}(D^- \to \mu^- \bar{\nu}_\mu)$ & $(3.77 \pm 0.18) \times 10^{-4}$ & \cite{HFLAV2019} \\
		$\mathcal{B}(D^- \to \tau^- \bar{\nu}_\tau)$ & $(1.20 \pm 0.27) \times 10^{-3}$ & \cite{HFLAV2019} \\
		$\mathcal{B}(K^- \to e^- \bar{\nu}_e)$ & $(1.582 \pm 0.007) \times 10^{-5}$ & \cite{PDG2020} \\
		$\mathcal{B}(K^- \to \mu^- \bar{\nu}_\mu)$ & $0.6356 \pm 0.0011$ & \cite{PDG2020} \\
		$\mathcal{B}(\tau^- \to K^- \bar{\nu}_\tau)$ & $(0.6986 \pm 0.0085) \times 10^{-2}$ & \cite{HFLAV2019} \\
		$\mathcal{B}(K^- \to \mu^- \bar{\nu}_\mu)/\mathcal{B}(\pi^- \to \mu^- \bar{\nu}_\mu)$ & $1.3367 \pm 0.0029$ & \cite{PDG2020} \\
		$\mathcal{B}(\tau^- \to K^- \bar{\nu}_\tau)/\mathcal{B}(\tau^- \to \pi^- \bar{\nu}_\tau)$ & $(6.438 \pm 0.094) \times 10^{-2}$ & \cite{HFLAV2019} \\
		$\mathcal{B}(B_s \to \mu^+ \mu^-)$ & $(2.9 \pm 0.7 \pm 0.2) \times 10^{-9}$ & \cite{CMS:2019bbr} \\
		$|V_{cd}|f_+^{D \to \pi}(0)$ & $0.1426 \pm 0.0018$ & \cite{HFLAV2019}\\
		$|V_{cs}|f_+^{D \to K}(0)$ & $0.7180 \pm 0.0033$ & \cite{HFLAV2019}\\
		$|\varepsilon_K|$ & $(2.228 \pm 0.011) \times 10^{-3}$ & \cite{PDG2020} \\
		$\Delta m_d$ & $(0.5065 \pm 0.0019)$ ps$^{-1}$ & \cite{HFLAV2019}\\
		$\Delta m_s$ & $(17.757 \pm 0.021)$ ps$^{-1}$ & \cite{HFLAV2019}\\
		sin~$2\beta$ & $0.71 \pm 0.09$ & \cite{HFLAV2019} \\
		$\phi_s$ & $-0.055 \pm 0.021$ & \cite{HFLAV2019}\\
		$\alpha$ & $(85.2^{+4.8}_{-4.3})^\circ$ & \cite{HFLAV2019} \\
		$\gamma$ & $(67 \pm 4)^\circ$ & \cite{LHCb:2020kho} \\
		\hline
		\hline
		$V_L$ & $0.995 \pm 0.021$ & \cite{PDG2020}\\
		$V_R$ & $[-0.11,0.16]$ & \cite{CMS:2020ezf}\\
		\hline
	\end{tabular}
	\caption{List of observables used for the CKM fit (\textit{Updated 2021}) in the SM framework. For the NP analysis we have not used the inclusive measurements of $|V_{ub}|$ and $|V_{cb}|$. All other inputs have been considered. Additionally, we have also considered the anomalous $Wtb$ couplings as listed in the last two rows.}
	\label{tab:CKM-updated-obs}
\end{table}

\begin{table}[h!!!]
	\centering
	\renewcommand{\arraystretch}{1.2}
	\begin{tabular}{|c|c|c|}
		\hline
		Input Parameters & Value & Reference \\
		\hline
		$f_+^{K \to \pi}(0)$ & $0.9706(27)$ & $N_f = 2+1+1$ \cite{FLAG:2019}\\
		$f_{K^\pm}/f_{\pi^\pm}$ & $1.1932(19)$ & $N_f = 2+1+1$\cite{FLAG:2019} \\
		$f_K$ & $155.7 \pm 0.13$ & $N_f = 2+1+1$ \cite{FLAG:2019}\\
		$f_+^{DK}(0)$ & $0.747 (19)$ & $N_f = 2+1+1$ \cite{FLAG:2019}\\
		$f_+^{D\pi}(0)$ & $0.666 (29)$ & $N_f = 2+1$ \cite{FLAG:2019}\\
		$f_{B_s}$ & $230.3(1.3)$ MeV & $N_f = 2+1+1$ \cite{FLAG:2019}\\
		$f_{B_s}/f_B$ & $1.209(0.005)$ & $N_f = 2+1+1$ \cite{FLAG:2019}\\
		$B_K$ & $0.7625(97)$ & $N_f = 2+1$ \cite{FLAG:2019}\\
		$f_{D_s}$ & $249.9(5)$ MeV & $N_f = 2+1+1$ \cite{FLAG:2019}\\
		$f_{D_s}/f_D$ & $1.1783(16)$ & $N_f = 2+1+1$ \cite{FLAG:2019}\\
		$\zeta(\Lambda_p \to p \mu^- \bar{\nu}_\mu)_{q^2 > 15}/\zeta(\Lambda_p \to \Lambda_c \mu^- \bar{\nu}_\mu)_{q^2 > 7}$ & $1.471 \pm 0.096 \pm 0.290$ & \cite{CKMFitter} \\
		$B_{B_s}$ & $1.327 \pm 0.016 \pm 0.030$ & \cite{FLAG:2019} \\
		$B_{B_s}/B_{B_d}$ & $1.007 \pm 0.013\pm 0.014$ & $N_f = 2$ \cite{FLAG:2019} \\
		$\bar{m}_c (m_c)$ & $1.2982\pm 0.0013\pm 0.0120$ GeV & \cite{CKMFitter}\\
		$\bar{m}_t (m_t)$ & $(165.26 \pm 0.11 \pm 0.30$ GeV & \cite{CKMFitter}\\
		$\eta_{tt}$ & $0.402 \pm 0 \pm 0.007$ & \cite{CKMFitter} \\
		$\eta_{ut}$ & $0.55 \pm 0 \pm 0.024$ & \cite{CKMFitter} \\
		$\eta_B (\bar{M}S)$ & $0.5510 \pm 0 \pm 0.0022$ & \cite{CKMFitter} \\
		\hline
	\end{tabular}
	\caption{List of aditional inputs for the CKM fit.}
	\label{tab:CKM-updated-inputs}
\end{table}

As we have mentioned in the previous section, the NP contributions to semileptonic ($P \to M\ell\nu_{\ell}$) and leptonic decays will impact the vertex factor, which is proportional to the square of the magnitude of the corresponding CKM element. Hence, we need to extract the parameters related to NP alongside the other Wolfenstein parameters. This means that we need to carry out a dedicated fit to all of these parameters using the machinery used by the CKMFitter group to fit only the CKM parameters.   

\begin{figure}[t]
	\centering
	{\includegraphics[scale=0.36]{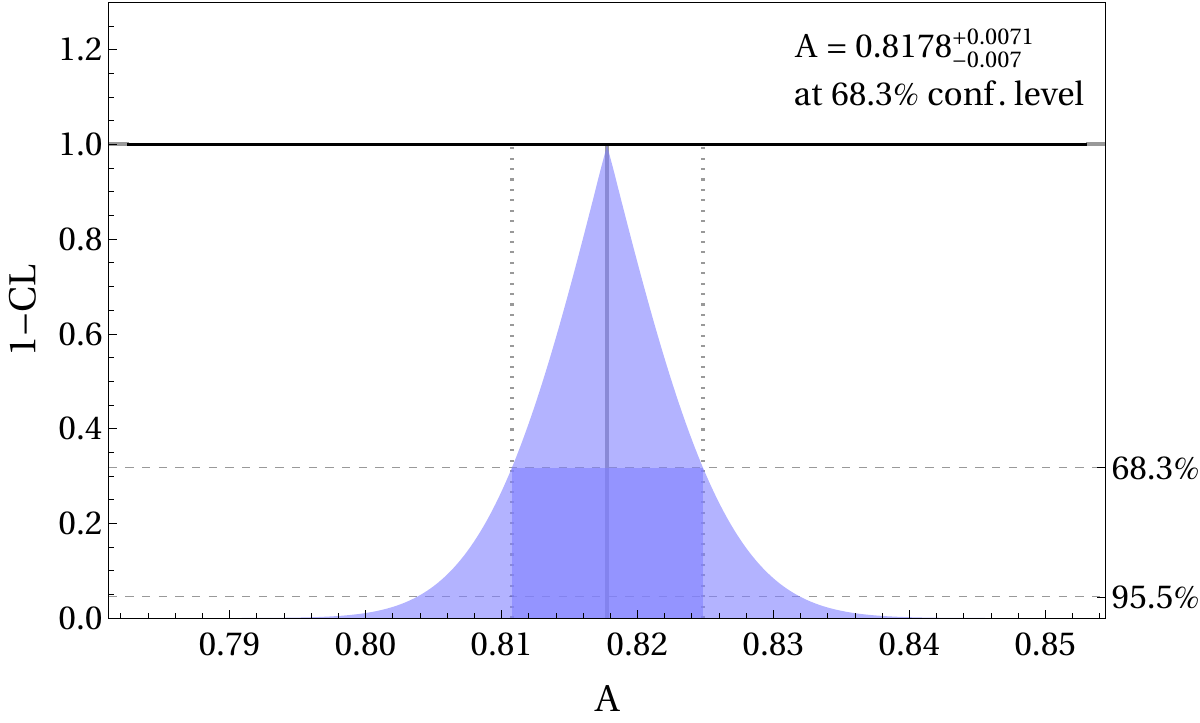}
		\label{fig:Inputs2021-PL-A}}~~
	{\includegraphics[scale=0.36]{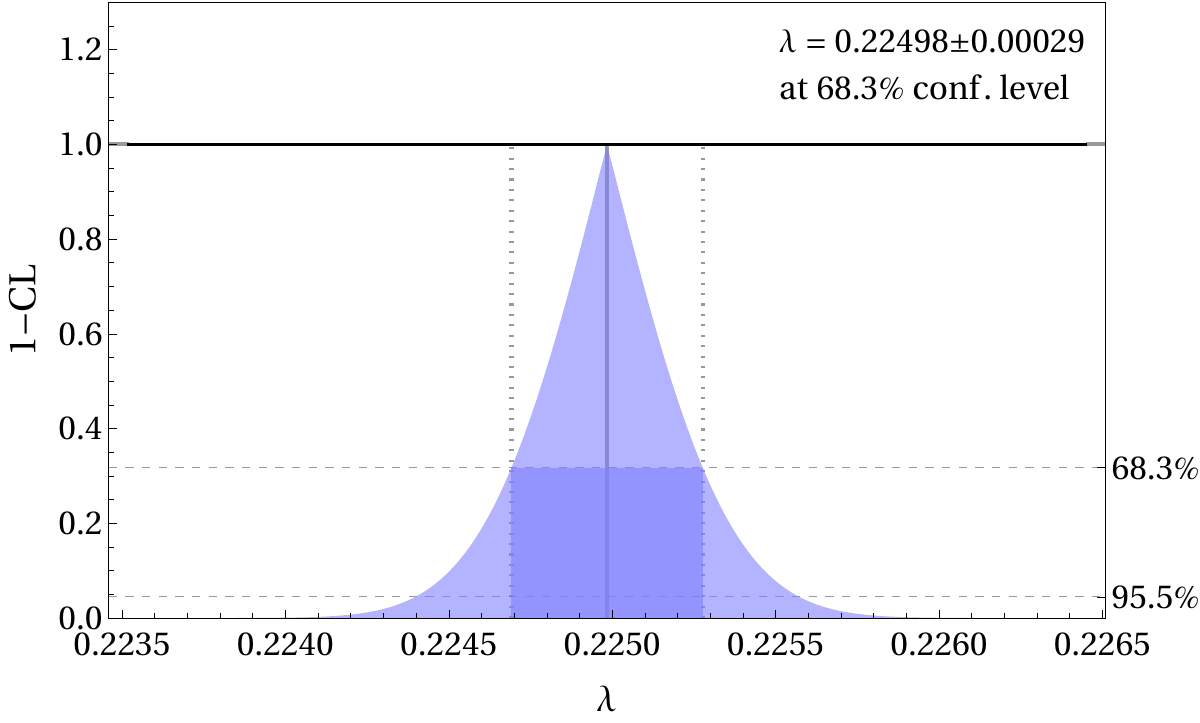}
		\label{fig:Inputs2021-PL-lam}}\\
	{\includegraphics[scale=0.36]{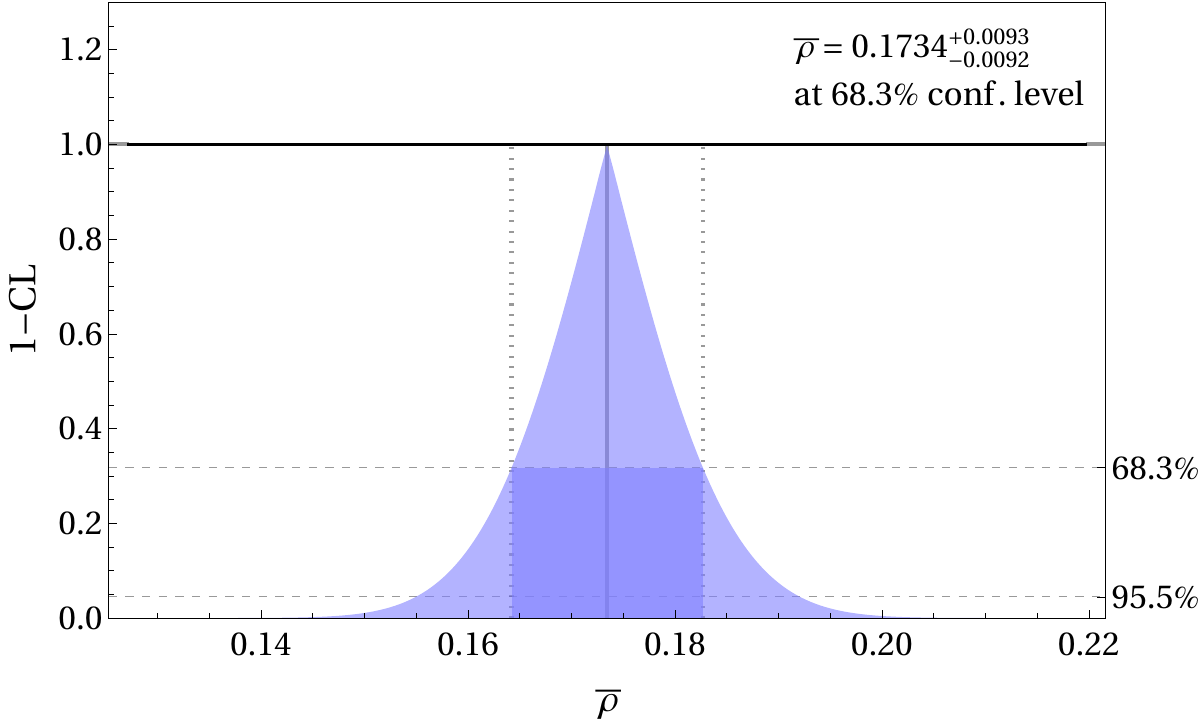}
		\label{fig:Inputs2021-PL-rhob}}~~
	{\includegraphics[scale=0.36]{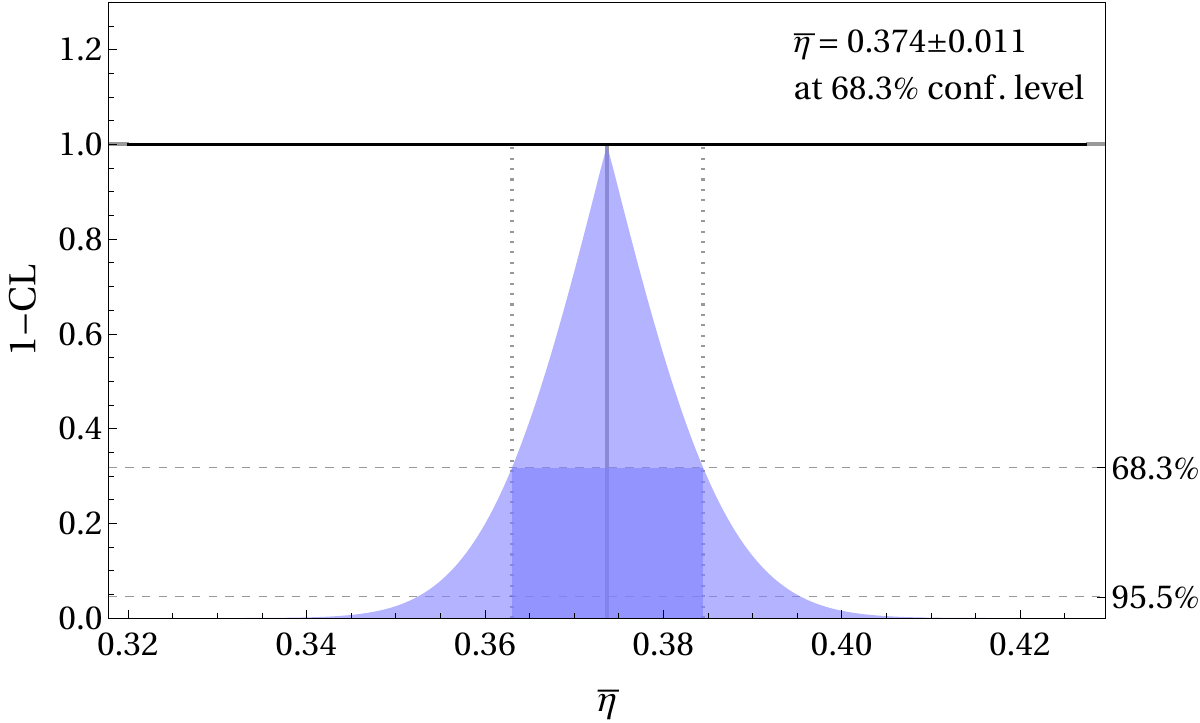}
		\label{fig:Inputs2021-PL-etab}}~~
	\caption{1D profile-likelihoods for the CKM Wolfenstein parameters $A,\lambda, \bar{\rho} ,\bar{\eta}$ for the global CKM 2021 Standard Model fit. The best fit estimates at $68.3\%$ confidence level are mentioned in each case.}
	\label{fig:Inputs2021-PL}
\end{figure}

To validate the code, we recreate the Summer'19 SM fit performed by the CKMFitter group using the same set of inputs and observables as mentioned in \cite{CKMFitter2019}. The details of the theoretical expressions for the observables can be found in \cite{HFLAV2019, Charles:2015gya, Charles:2011va, Charles:2004jd}. We report our fit results in Table.~\ref{tab:CKM-fit} and compare them to the CKMFitter 2019 results. They are consistent with each other within $1\sigma$ confidence interval (CI). We go a step further and use some recent updates for the CKM observables as listed in  Table.~\ref{tab:CKM-updated-obs} and redo the fit in this ``\textit{Updated 2021}'' scenario. 
This is the most updated global fit results after CKMFitter 2019. The other relevant inputs are provided in Table.~\ref{tab:CKM-updated-inputs}. Note that the fit results for all the four parameters are consistent with 2019 results within $1\sigma$ CI. However, the fit values for $\bar \rho$ and $\bar \eta$ are slightly higher than earlier. The best fit points for $\bar \rho$ has increased by 8\% while that for $\bar \eta$ by about 5\%. Primarily, these shifts are due to changes in the inputs of $\alpha$, $\gamma$ and $\sin 2\beta$ which have been updated from the previous 2019 inputs. Fig.~\ref{fig:Inputs2021-PL} shows the single parameter profile-likelihoods for the global CKM fit with the most updated inputs and observables. These are the most updated best fit estimates for the CKM parameters. 

\subsection{CKM Fit including new physics}
As mentioned earlier, due to the presence of the WC corresponding to the $V+A$ operator $\mathcal{O}_{V_2}$, the decay distribution of $P \to M^* \ell \nu_\ell$ decays will be modified unlike the alteration of the vertex CKM factor in case of the $P \to M \ell \nu_\ell$ and $P \to \ell \nu_\ell$ decays. Hence, in order to perform the fit for the NP scenarios, we consider both the CKM observables listed in Table.~\ref{tab:CKM-updated-obs} as well as the list of $B\to D^* \ell \bar{\nu}_\ell$ data mentioned in the previous subsection. However, we do not consider the inclusive determinations of $|V_{ub}|$ and $|V_{cb}|$ for the NP fit as mentioned earlier. Additionally, we also consider the $Wtb$ anomalous couplings as observables which are significantly affected by heavy BSM physics like the present model. The most general $Wtb$ vertex is expressed as
\beq
\mathcal{L}_{Wtb} = -\frac{g}{\sqrt{2}}\bar{b}\gamma^\mu (V_L P_L + V_R P_R) t W^- -\frac{g}{\sqrt{2}}\bar{b}\frac{i \sigma^{\mu\nu}}{M_W}(g_L P_L + g_R P_R)t W^- + h.c.
\eeq
In the SM, $V_L = V_{tb} \simeq 1$ while the other anomalous couplings $V_R,g_L,g_R$ are equal to zero. Limits have been set on such couplings by analyzing CMS and ATLAS data on helicity fractions, single top production cross section and forward-backward asymmetries \cite{Deliot:2017byp, Deliot:2018jts, CMS:2020ezf}. Since our model will have contributions to these couplings which are much enhanced due to the top quark mass, we use them as inputs as listed in Table.~\ref{tab:CKM-updated-obs}. Note that, the NP contribution to the tensor operator $\sigma^{\mu\nu}$ is much suppressed and therefore we do not consider the anomalous couplings $g_L$ and $g_R$ in our analysis. 

\begin{table}[htp!!]
\small
\centering
\renewcommand{\arraystretch}{1.2}
\begin{tabular}{|c|c|c|c|c|}
\hline
\multirow{2}{*}{Scale $\Lambda$ (TeV)} & \multicolumn{2}{|c|}{Fit Quality} & \multirow{2}{*}{Parameter} & \multirow{2}{*}{Fit Result}  \\
\cline{2-3} 
& $\chi^2/dof$ & p-Value & &  \\
\hline
\multirow{2}{*}{$1.0$} & \multirow{2}{*}{$42.57/44$}& \multirow{2}{*}{$53.31\%$} & $C_T$ (GeV$^{-2}$) & $0.306 (85)$  \\
& & & $|V_{cb}|$ & $40.82 (93) \times 10^{-3} $ \\
& & & $a_{0}^f$ & $0.0122 (1)$ \\
& & & $a_{1}^f$ & $0.0181 (96)$ \\
& & & $a_{2}^f$ & $-0.268 (210)$ \\
& & & $a_{0}^g$ & $0.0333 (11)$ \\
& & & $a_{1}^g$ & $-0.099 (64)$ \\
& & & $a_{2}^g$ & $-0.058 (148)$ \\
& & & $a_{1}^{\mathcal{F}_1}$ & $-0.0003 (17)$ \\
& & & $a_{0}^{\mathcal{F}_2}$ & $0.0513 (12)$ \\
& & & $a_{1}^{\mathcal{F}_2}$ & $-0.185 (60)$ \\
& & & $a_{2}^{\mathcal{F}_2}$ & $0.981 (921)$ \\
\hline
\multirow{2}{*}{$2.0$} & \multirow{2}{*}{$42.57/44$}& \multirow{2}{*}{$53.31\%$} & $C_T$ (GeV$^{-2}$) & $0.272 (75)$ \\
& & & $|V_{cb}|$ & $40.82 (93) \times 10^{-3} $ \\
\hline
\multirow{2}{*}{$5.0$} & \multirow{2}{*}{$42.57/44$}& \multirow{2}{*}{$53.31\%$} & $C_T$ (GeV$^{-2}$) & $0.237 (66)$ \\
& & & $|V_{cb}|$ & $40.82 (93) \times 10^{-3} $ \\
\hline
\end{tabular}
\caption{Fit result for $|V_{cb}|$ and $C_T$ (GeV$^{-2}$) from the frequentist analysis for different NP scenario with the same $B \to D^* \ell \bar{\nu}_\ell$ dataset as Table.~\ref{tab:BDstlnu-SM}. We have shown the fit results for the BGL coefficients only for $\Lambda = 1$ TeV. For $\Lambda = 2$ and $5$ TeV, the results are identical.  }
\label{tab:BDstlnu-NP}
\end{table}

\begin{table}[h!!!]
\renewcommand{\arraystretch}{1.5}
\centering
\resizebox{\textwidth}{!}{
\begin{tabular}{|cc|c|c|c|c|c|c|c|}
\hline
\multicolumn{2}{|c|}{\multirow{2}{*}{$\text{Case}$}}  &  \multirow{2}{*}{$\chi^2\text{/dof}$}  &  \multirow{2}{*}{$\text{p-Value ($\%$)}$} &  \multicolumn{5}{c|}{$\text{Fit Result}$} \\
\cline{5-9}
 &&  &  &  $C_T$ (GeV$^{-2}$)  &  $\text{A}$  &  $\lambda$ &  $\bar{\rho}$  &  $\bar{\eta}$  \\
 \hline
\multirow{3}{*}{$\text{Scale} \Bigg \lbrace $ } & $1 \text{ TeV}$  &  $87.49/70$  &  $7.69$  &  $0.00003 \pm 0.00013$  & $0.799806 \pm 0.007691$  &  $0.224982 \pm 0.000293$  &  $0.176546 \pm 0.009690$  &  $0.386274 \pm 0.011863$  \\
& $2
\text{ TeV}$  &  $87.49/70$  &  $7.69$  &  $0.00002 \pm 0.00009$  & $0.799806 \pm 0.007691$  &  $0.224982 \pm 0.000293$  &  $0.176545 \pm 0.009689$  &  $0.386274 \pm 0.011863$  \\
& $5
\text{ TeV}$  &  $87.49/70$  &  $7.69$  &  $0.000015 \pm 0.000066$  & $0.799808 \pm 0.007691$  &  $0.224982 \pm 0.000293$  &  $0.176544 \pm 0.009690$  &  $0.386273 \pm 0.011863$  \\
\hline
\end{tabular}}
\caption{Fit Results for the Wolfenstein parameters and $C_T$ with NP. For the NP analyses, we have shown the results for three benchmark values of the cutoff scale, $\Lambda = (1.0, 2.0, 5.0)$ TeV. The corresponding results for the BGL coefficients are given in Table.~\ref{tab:bglfit} in the appendix.}
 \label{tab:comb-NP-rslt}
\end{table}

To begin with, we present the fit results corresponding to the analysis of $B\to D^* \ell \bar{\nu}_\ell$ alone in Table.~\ref{tab:BDstlnu-NP}. We fit $C_T = C_s^2 + C_p^2$ along with $|V_{cb}|$ and the BGL coefficients for different values of $\Lambda$ in between 1 to 10. However, we have presented the results only for three different values of the cut-off scale $\Lambda$, for example, $\Lambda$= 1, 2 and 5 TeV, respectively. In the cases we have studied for different values of $\Lambda$, the fitted values for the BGL coefficients are identical, and we present them only for $\Lambda = 1$ TeV. Note that because of the new contribution in the decay rate distribution, there is a small shift ($\approx 5.5\%$) in the best-fit values of $|V_{cb}|$. However, the fitted values are consistent within 1-$\sigma$ CI with the one obtained without any NP (Table.~\ref{tab:BDstlnu-SM}). We obtain a non-zero solution for $C_T$, which is allowed due to a small discrepancy between the Fermilab-MILC estimates and the measurement of the decay rates, which can be seen in Fig. 8 of ref. \cite{FermilabLattice:2021cdg}. 

As a next step, we include the data on $B\to D^* \ell \bar{\nu}_\ell$ alongside all the other data used in the CKM fit. The presence of a new contribution in $P\to M\ell\nu_{\ell}$ and $P \to \ell\nu_{\ell}$ decays modifies the CKM element to $|V_{ij}^{\prime}| = |V_{ij}(1 \pm C_{V_2})|$ (with $C_{V_1} =0$). In such cases, the measured values of the elements should be considered to be $|V_{ij}^{\prime}|$ while $|V_{ij}|$ will be parametrized in terms of $A$, $\lambda$, $\bar{\rho}$ and $\bar{\eta}$. In the expansion of $V_{ij}$ we consider terms up to order $\lambda^8$. The fit results of the corresponding frequentist analysis are presented in Table.~\ref{tab:comb-NP-rslt} \footnote{ In Table.~\ref{tab:comb-NP-rslt-2}, we list the fit results for the Wolfenstein parameters including NP in presence of the Belle 2017 data and find no significant changes in the fit results compared to those listed in Table.~\ref{tab:comb-NP-rslt}.}.  Note that in the presence of NP, the fitted values of $A$, $\lambda$, $\bar{\rho}$ and $\bar{\eta}$ remain practically unchanged. For both the values of $\Lambda$ the allowed ranges of $C_T$ are consistent with zero and very tightly constrained. The negative values of $C_T$ could be accomodated by introducing phases in $C_s$ and $C_p$, for example, by the following replacements: $C_s \to e^{i \pi/2} C_s = i C_s$ and $C_p \to e^{i \pi/2} C_p = i C_p$\footnote{In principle, one can consider $C_s$ and $C_p$ to be complex with the respective phases as unknowns which can be constrained from the data on mixing and eletric dipole moments etc. We did not explore that possibility which we will do in a future work. Furthermore, it is to be noted that our NP scenario has negligible impacts on $K-{\bar K}$ or $B_q -{\bar B_q}$ (q =d,s) mixing.}.     

\begin{table}[t]
	\begin{tabular}{|c|c|c|c|c|}
		\hline
		\textbf{Parameters} & \multirow{3}{*}{\textbf{Without NP}} & \multicolumn{3}{c|}{\textbf{In scenarios with NP}}\\
		\cline{3-5}
		  &   &  $\text{$\Lambda $=1 TeV}$  & $\text{$\Lambda $=2 TeV}$ & $\text{$\Lambda $=5 TeV}$ \\ 
		  \hline 
$\text{A}$  &  $0.79925_{-0.00757}^{+0.00766}$  &  $0.79922_{-0.00753}^{+0.00767}$  &  $0.79922_{-0.00753}^{+0.00765}$  &  $0.79943_{-0.00759}^{+0.00769}$  \\
  $\lambda$  &  $\text{0.224979$\pm $0.000293}$  &  $0.224979 \pm 0.000293$  &  $0.22498_{-0.000292}^{+0.000294}$  &  $\text{0.224979$\pm $0.000292}$  \\
  $\bar{\rho }$  &  $0.17657_{-0.00969}^{+0.00971}$  &  $0.1765_{-0.00962}^{+0.00959}$  &  $0.17658_ {-0.00963}^{+0.0097}$  &  $0.17628_{-0.0096}^{+0.00977}$  \\
  $\bar{\eta }$  &  $0.3867_{-0.0118}^{+0.0119}$  &  $0.3866_{-0.0117}^{+0.0118}$  &  $0.3865_{-0.0118}^{+0.0119}$  &  $0.3862_{-0.0118}^{+0.0119}$  \\ \hline
		$C_T$ (GeV$^{-2}$)  &  N.A.  &  $0.0000297_{-0001263}^{+0001260}$  &  $0.0000214 \pm 0.0000906$    & $0.0000156_{-0.000065}^{+0.0000659}$\\\hline
	\end{tabular}
	\caption{The extracted values of the Wolfenstein parameters in the bayesian fit with and without the contributions from NP. We have considered the NP scale $\Lambda$ to be 1, 2 and 5 TeV in the NP scenarios, respectively. The numbers correspond to the medians and $1\sigma$ quantiles of the respective distributions for the CKM parameters. The corresponding results for the BGL coefficients are given in Table.~\ref{tab:bglfit} in the appendix.}
	\label{tab:CKM_par}
\end{table}

\begin{table}[htbp]
	\begin{tabular}{|c|c|c|c|c|}
		\hline
		\textbf{CKM elements} & \multirow{2}{*}{\textbf{Without NP}} & \multicolumn{3}{c|}{\textbf{In scenarios with NP}}\\
		\cline{3-5}
		$$  &  $$  &  $\text{$\Lambda $=1 TeV}$  &  $\text{$\Lambda $=2 TeV}$  &  $\text{$\Lambda $=5 TeV}$\\\hline
		$\left|V_{\text{ud}}\right|$  &  $\text{0.974355$\pm $0.000068}$ &  $\text{0.974356$\pm $0.000068}$  &  $0.974355_{-0.000068}^{+0.000067}$  &  $\text{0.974356$\pm $0.000067}$  \\
		$\left|V_{\text{us}}\right|$  &  $\text{0.22498$\pm $0.00029}$ &  $\text{0.22498$\pm $0.00029}$  &  $\text{0.22498$\pm $0.00029}$ &  $\text{0.22498$\pm $0.00029}$  \\
		$\left|V_{\text{ub}}\right|$  &  $\text{0.00397$\pm $0.00011}$ &  $\text{0.00397$\pm $0.00011}$  &  $\text{0.00397$\pm $0.00011}$ &  $\text{0.00396$\pm $0.00011}$  \\
		$\left|V_{\text{cd}}\right|$  &  $\text{0.22486$\pm $0.00029}$ &  $\text{0.22486$\pm $0.00029}$  &  $\text{0.22486$\pm $0.00029}$ &  $\text{0.22486$\pm $0.00029}$  \\
		$\left|V_{\text{vs}}\right|$  &  $\text{0.97351$\pm $0.00007}$ &  $\text{0.973509$\pm $0.00007}$  &  $\text{0.973509$\pm $0.00007}$  &  $0.973509_{-0.00007}^{+0.000069}$  \\
		$\left|V_{\text{cb}}\right|$  &  $0.04045_{-0.00037}^{+0.00038}$  &  $\text{0.04045$\pm $0.00037}$  &  $\text{0.04045$\pm $0.00037}$ &  $0.04046_{-0.00037}^{+0.00038}$  \\
		$\left|V_{\text{td}}\right|$  &  $\text{0.00828$\pm $0.0001}$ &  $\text{0.00828$\pm $0.0001}$  &  $\text{0.00828$\pm $0.0001}$ &  $\text{0.00828$\pm $0.0001}$  \\
		$\left|V_{\text{ts}}\right|$  &  $0.0398_{-0.00036}^{+0.00037}$ &  $\text{0.03979$\pm $0.00036}$  &  $\text{0.03979$\pm $0.00036}$ &  $0.0398_{-0.00036}^{+0.00037}$  \\
		$\left|V_{\text{tb}}\right|$  &  $\text{0.999174$\pm $0.000015}$ &  $\text{0.999174$\pm $0.000015}$  &  $\text{0.999174$\pm $0.000015}$  &  $\text{0.999173$\pm $0.000015}$\\
		\hline
	\end{tabular}
	\caption{The extracted values of the CKM elements from the fit results given in Table.~\ref{tab:CKM_par}) in the different scenarios with and without the NP. These estimates have been obtained from the bayesian posteriors of the respective runs for the SM and NP scenarios with the NP scale $\Lambda$ taken to be 1, 2 and 5 TeV. The numbers correspond to the medians and $1\sigma$ quantiles of the respective distributions for the CKM elements. It is evident that the inclusion of NP has negligible effect on these elements.}	
	\label{tab:CKM}
\end{table}

\begin{table}[htbp]
	\centering
	\begin{tabular}{|c|c|c|c|c|}
		\hline
		\text{Observable}  & SM & \multicolumn{3}{c|}{\textbf{In scenarios with NP}}\\
		\cline{3-5}
		&  &   $\text{1 TeV}$  &  $\text{2 TeV}$ &  $\text{5 TeV}$  \\ 
		\hline 
		\multirow{2}{*}{$ R(D^*) \Bigg\{$}	\text{Frequentist} & $\text{0.2579$\pm $0.0034}$ & $\text{0.2579$\pm $0.0034}$ & $\text{0.2579$\pm $0.0034}$ & $\text{0.2579$\pm $0.0034}$ \\
		~~~~~~~~ \text{Bayesian} & $0.2586_{-0.0030}^{+0.0031}$ & $0.2586_{-0.0030}^{+0.0031}$ &  $0.2586_{-0.0030}^{+0.0031}$ & $0.2586_{-0.0030}^{+0.0031}$ \\
		\hline 
	\end{tabular}
	\caption{$R(D^*)$ estimates for the SM and the three NP scenarios with scales 1, 2 and 5 TeV. The Bayesian estimates correspond to the median and 1$\sigma$ quantiles for the respective distributions for $R(D^*)$.}
	\label{tab:RDst}
\end{table}

\begin{figure}[t]
	\centering
	\includegraphics[scale=0.4]{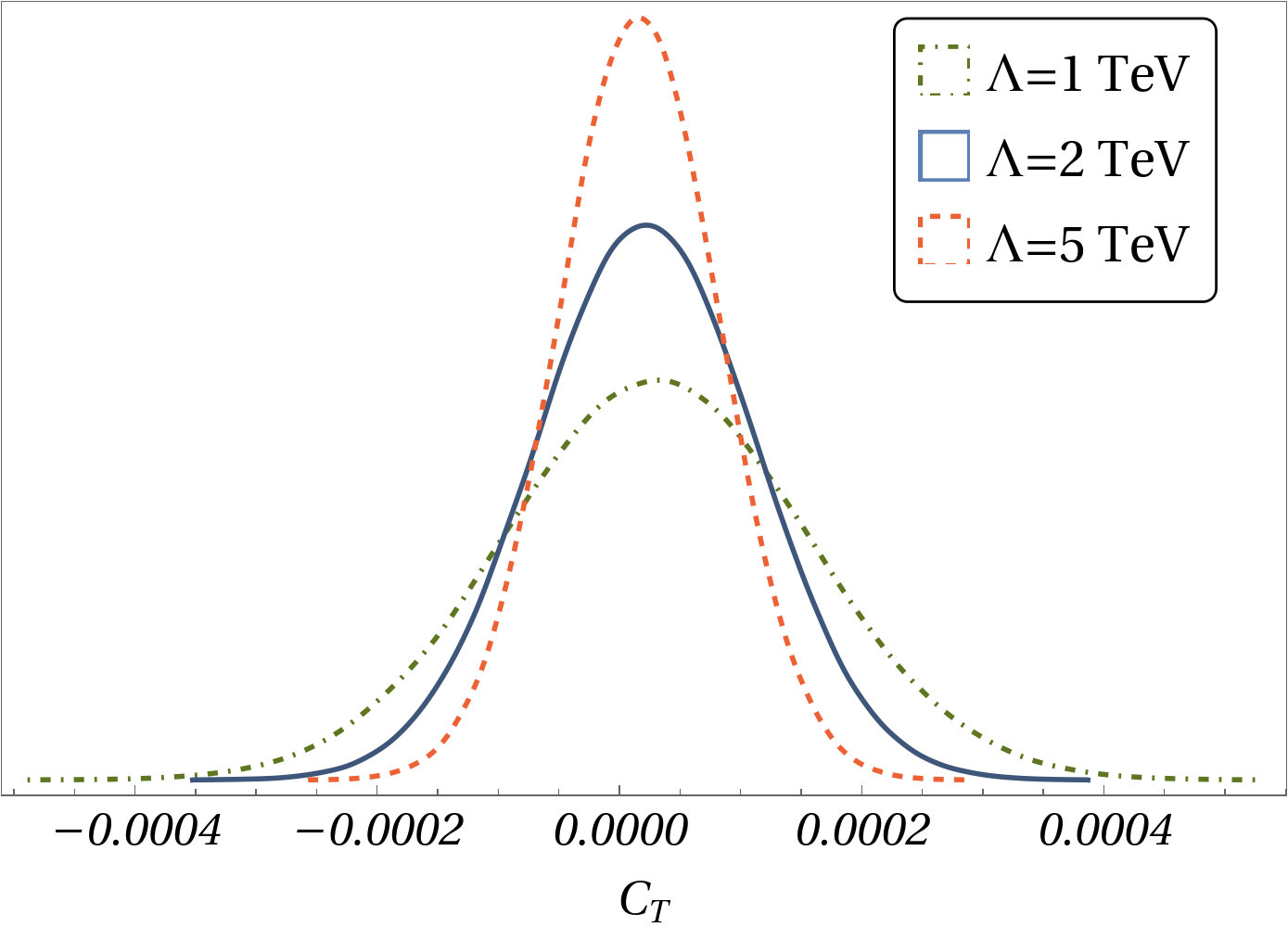}
	\caption{1D posteriors for the NP coupling $C_T$ corresponding to NP scale $\lambda$ taken to be 1 TeV, 2 TeV, and 5 TeV. 
	}
	\label{fig:NP_plt}
\end{figure}

\begin{figure}[t]
	\begin{center}
		\includegraphics[scale=0.25]{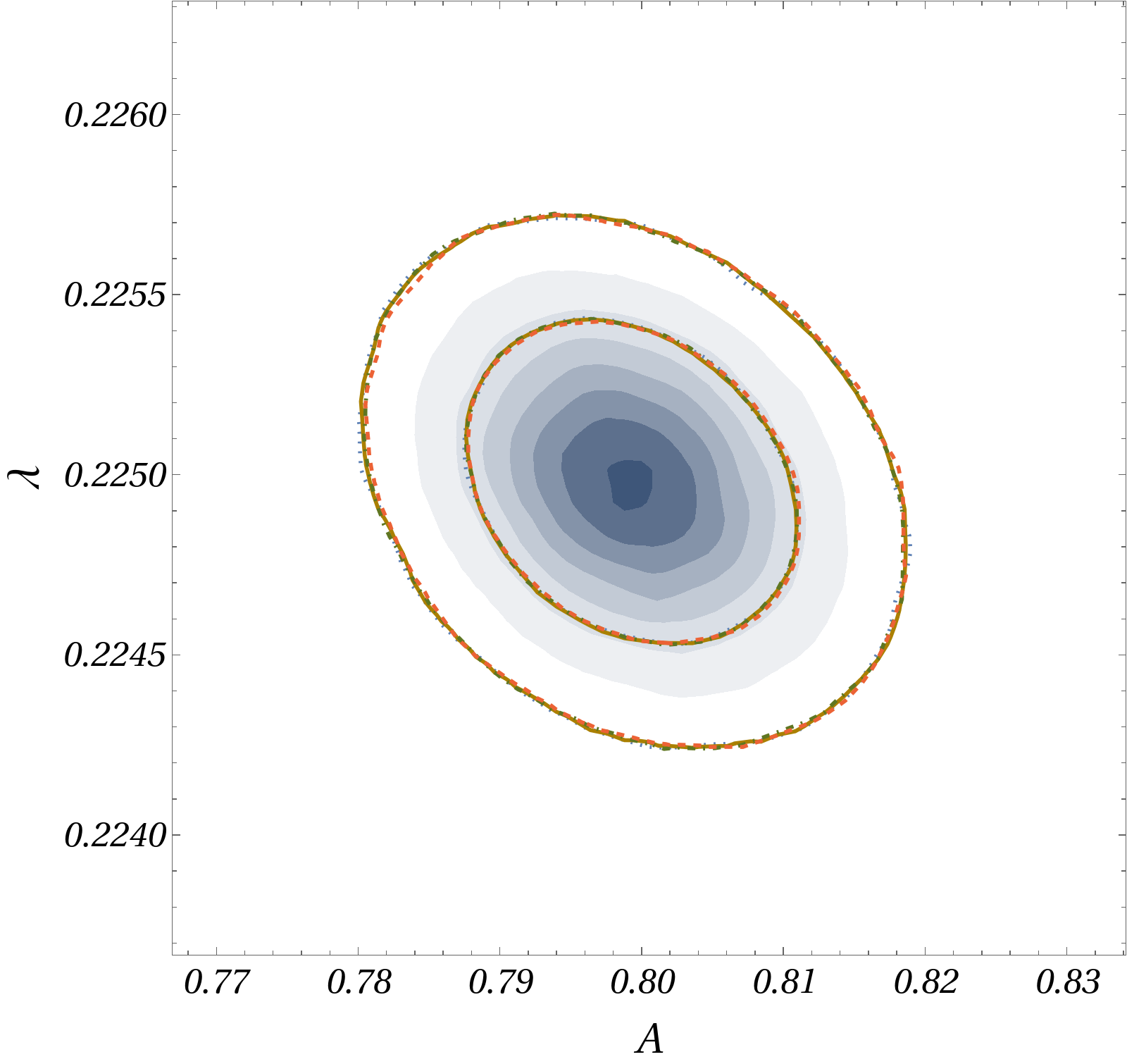}\label{fig:corAlam}~~
		\includegraphics[scale=0.25]{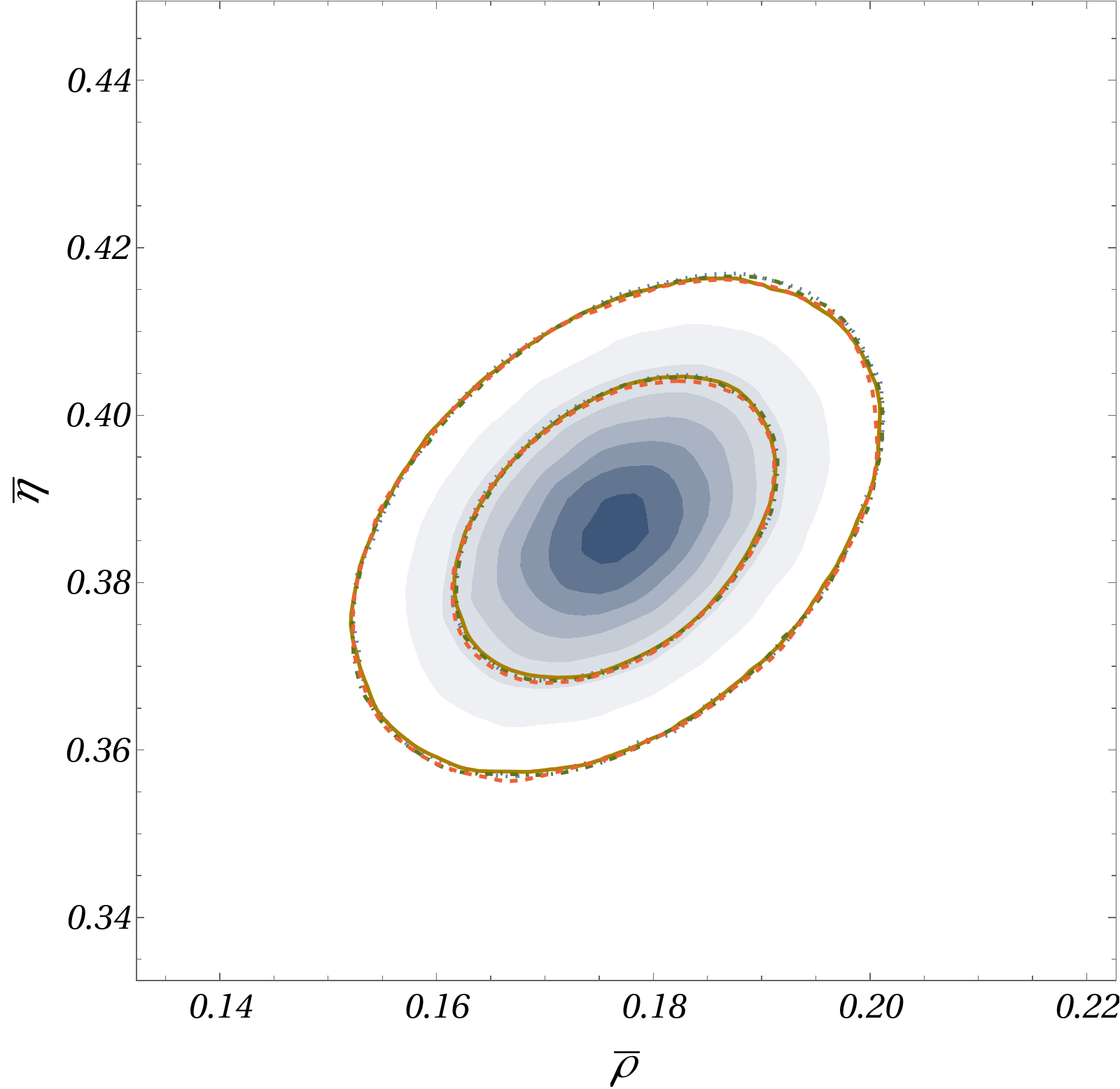}\label{fig:corrhoeta}~~
		\includegraphics[scale=0.25]{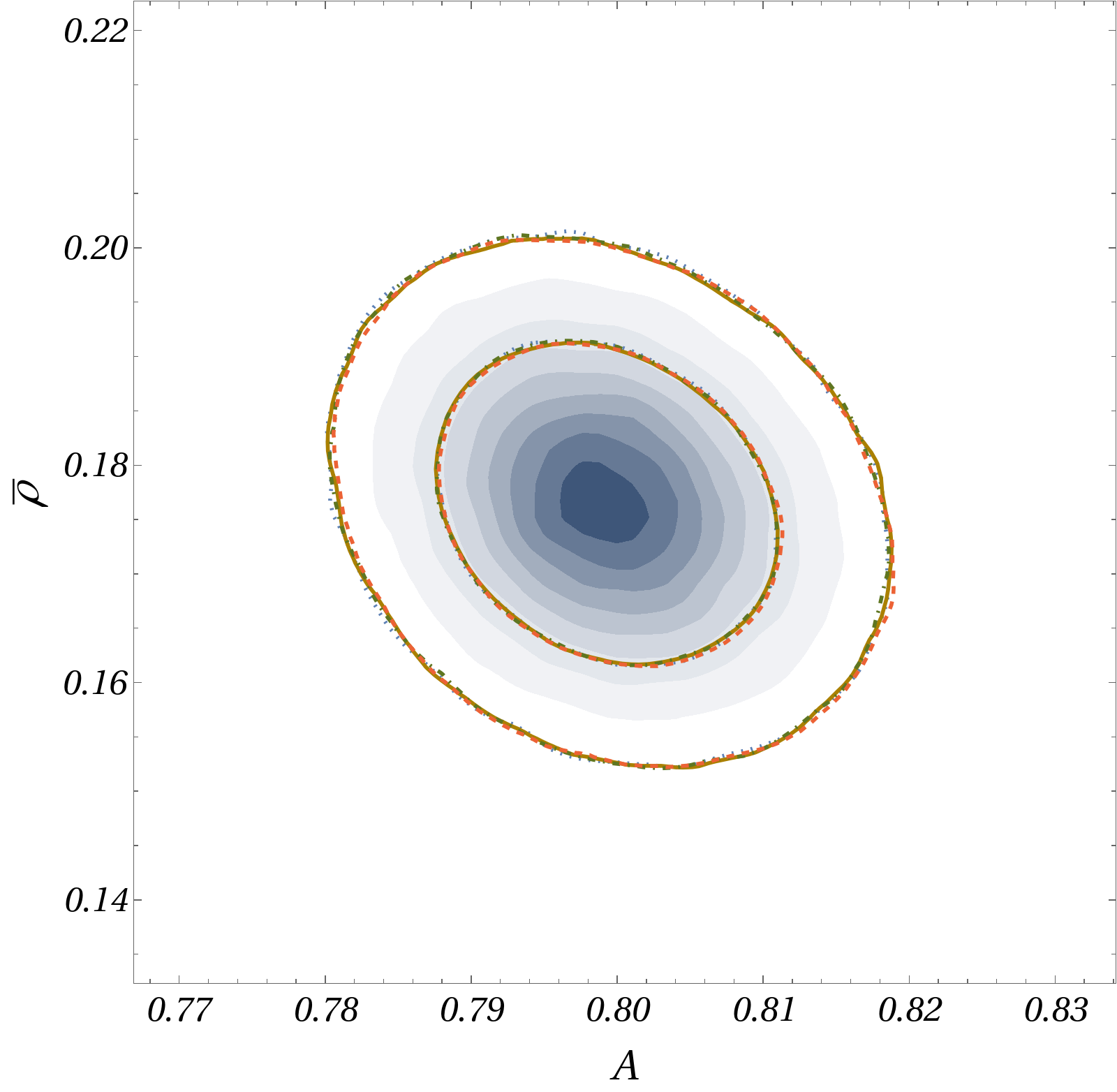}\label{fig:corArho}\\
		\includegraphics[scale=0.25]{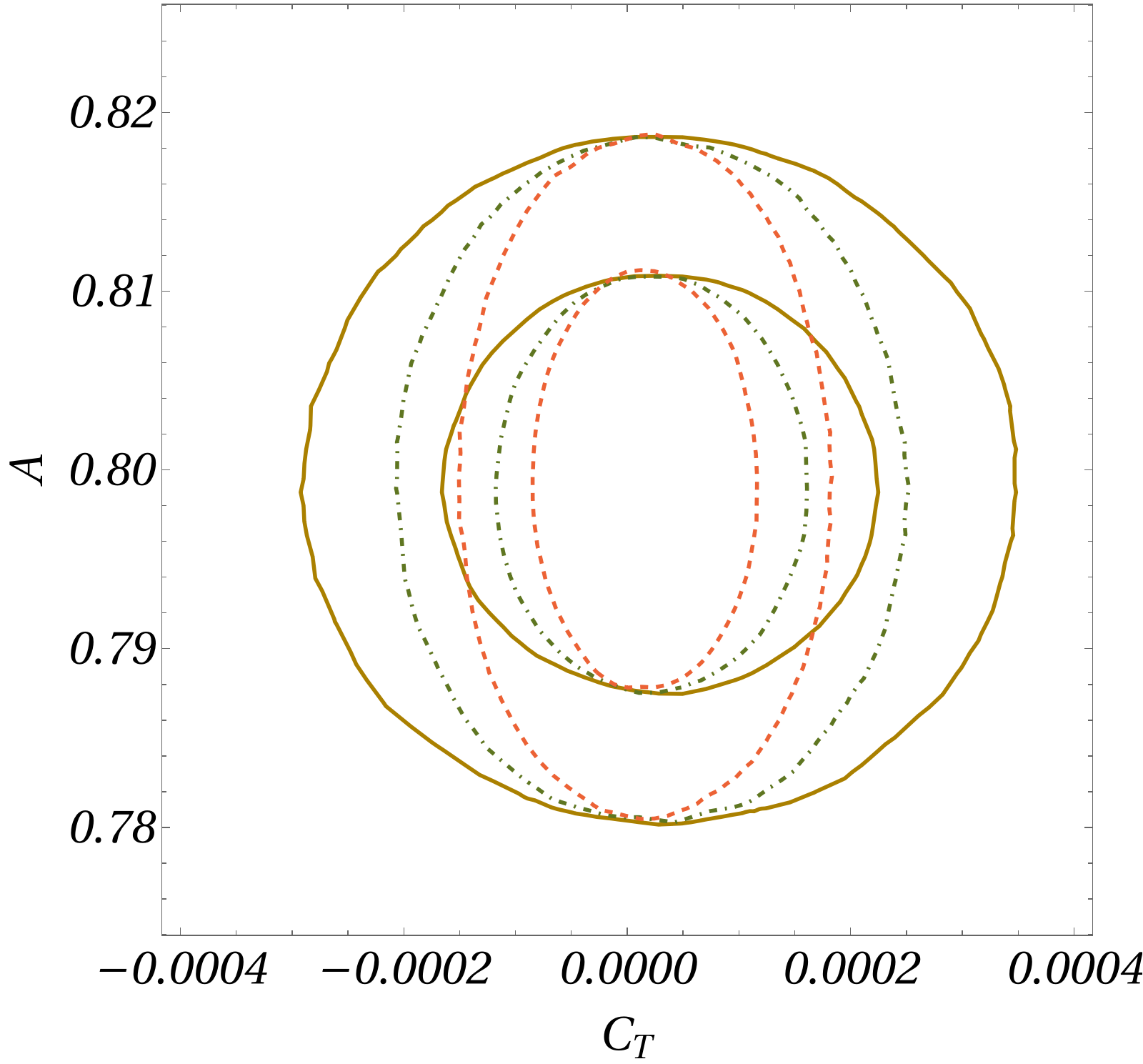}\label{fig:corACT}~~
		\includegraphics[scale=0.25]{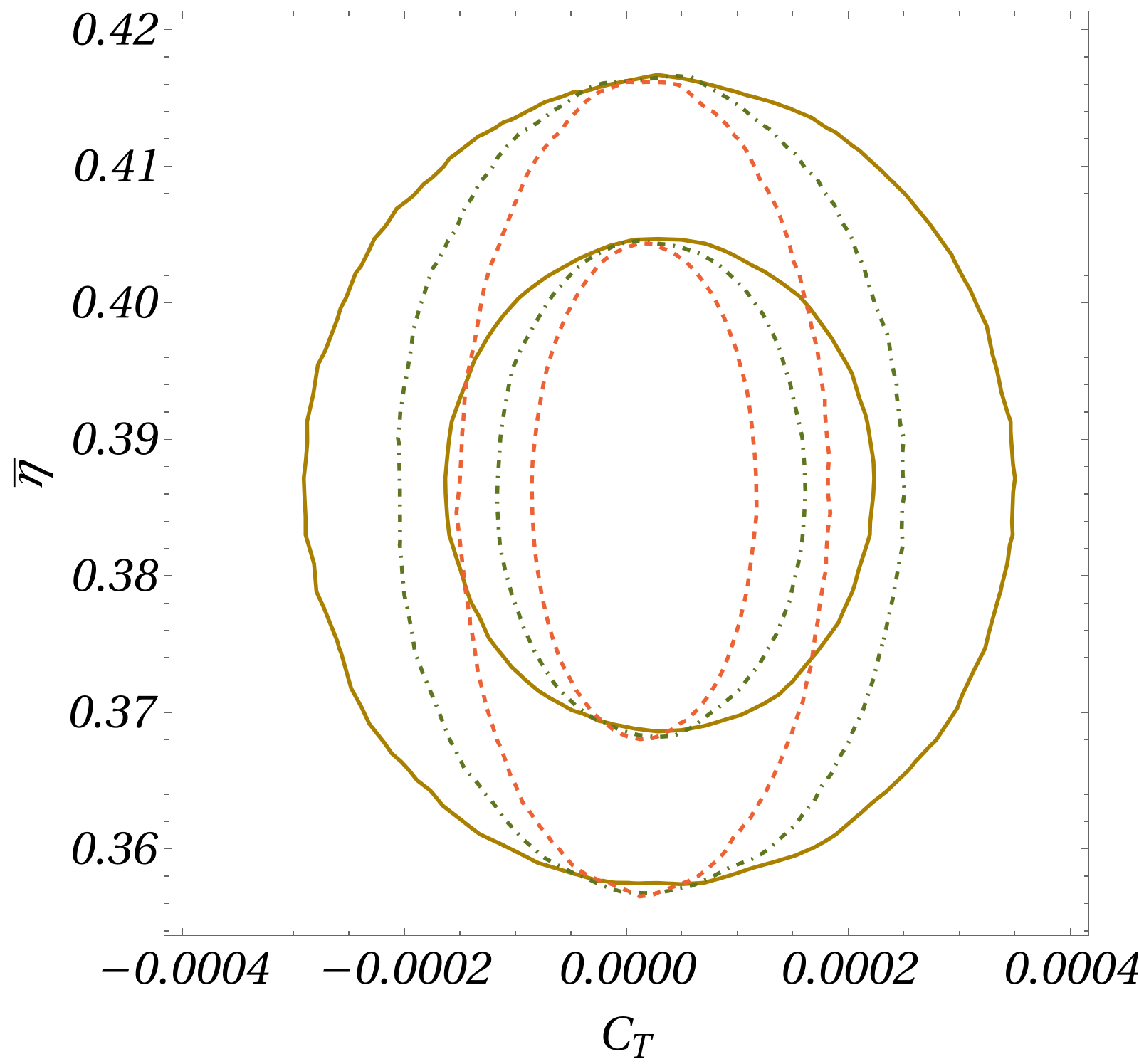}\label{fig:coretabCT}~~
		\includegraphics[scale=0.25]{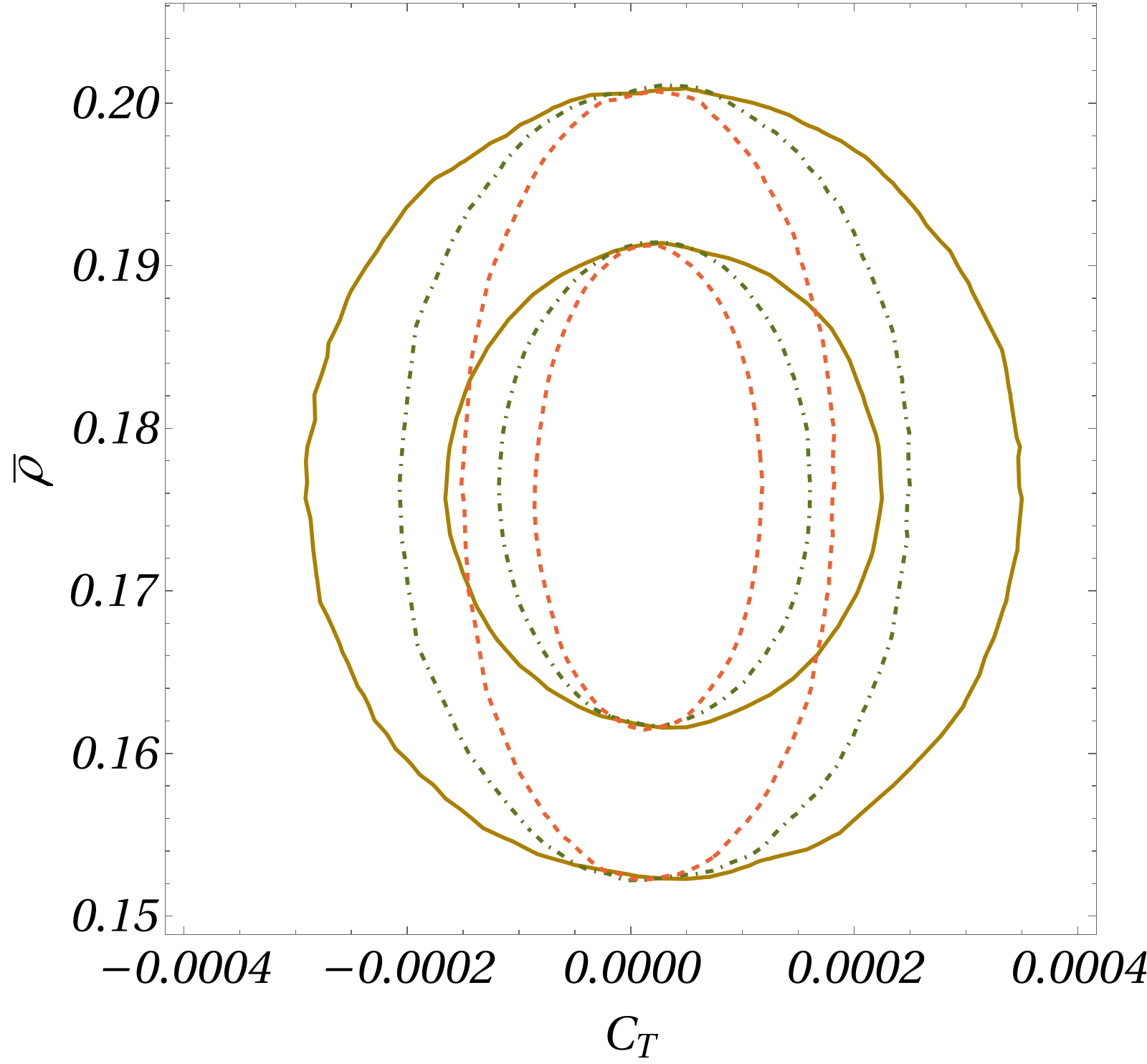}\label{fig:corrhobCT}\\
		\includegraphics[scale=0.5]{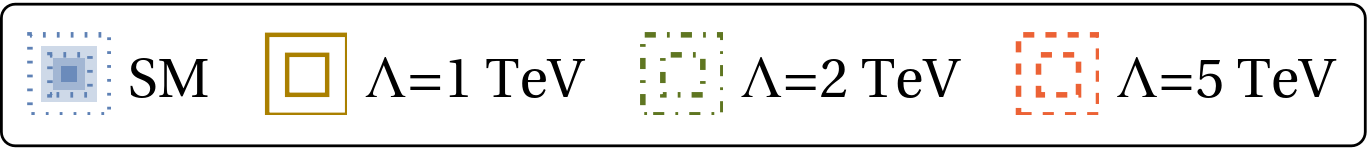}\label{fig:leg}
		\caption{2D correlation plots for the Wolfenstein parameters. We show the correlations between $A$-$\lambda$, $\bar{\rho}$-$\bar{\eta}$ and $A$-$ \bar{\rho}$ for the scenario without NP and the two NP cases with the NP scale $\Lambda$ taken to be 1, 2 and 5 TeV respectively. The smaller and larger concentric ellipses represent the 1 and 2$\sigma$ regions and have been displayed for the SM and all the NP cases. The shaded contours represent probability densities and have been provided only for the SM case. The blue (dotted) ellipses represent the SM while the brown (solid), green (dotdashed) and red (dashed) ellipses represent the NP cases with $\Lambda=$ 1, 2 and 5 TeV respectively.}
		\label{fig:corwolfp}
	\end{center}
\end{figure}

In the Bayesian view of subjective probability, all unknown parameters are treated as uncertain and thus should be described in terms of their underlying probability distributions.
In addition to the frequentist analysis, we also carry out a Bayesian fit for the Wolfenstein parameters with and without the contribution from the NP cases discussed above. The results of the bayesian fit are given in Table.~\ref{tab:CKM_par}. Note that the fit values of the Wolfenstein parameters are highly consistent in all the scenarios with and without the NP. All other observations are similar to the ones obtained in the frequentist analysis. In order to provide numerical estimates, we present the median and the corresponding 1$\sigma$ quantiles for the posteriors of the respective parameters. In the presence of the NP, the best fit points of all the Wolfenstein parameters are almost unchanged. The fitted values for $C_T$ are given in Table.~\ref{tab:CKM_par}. The corresponding $1D$ posterior has been shown in Fig.~\ref{fig:NP_plt}. In accordance to our expectations, $C_T$ is consistent with zero, and we obtain tight constraints on it which become more stringent with the increasing values of the cut-off scale. The overall observations remain similar to those obtained from the frequentist analysis. The posteriors for the Wolfenstein parameters: $A$, $\lambda$, $\bar{\rho}$ and $\bar{\eta}$ are understandably Gaussian. We refrain from showing the corresponding posteriors for all of the fit and nuisance parameters here. For the Bayesian analysis related to the BGL parameters, we provide the corresponding information consisting of the 1-D posteriors, 2-D correlation plots, and the corresponding numerical estimates as a triplot (Fig.~\ref{fig:triplot_BGL}) in section~\ref{BGL}.

In Fig.~\ref{fig:corwolfp},  we provide 2-D correlation plots between the CKM parameters $A-\lambda$, $\bar{\rho}-\bar{\eta}$ and $A-\bar{\rho}$. We also display the correlation of the NP parameter $C_T$ with $A$, $\bar \eta$ and $\bar{\rho}$. Note that in the presence of a new contribution the correlations between $A$, $\lambda$ $\bar{\eta}$ and $\bar\rho$ do not change. Also, for all values of the scale $\Lambda$, $C_T$ has negligible correlations with the Wofenstein parameters. In addition, we have checked that $C_T$ has a negligible correlation with $\lambda$ for all cases. The numerical values of these correlations are presented in the appendix. In the absence of any NP contributions, the numerical values of the correlations are given in Table.~\ref{tab:corrwolfSM}, \ref{tab:corrwolf1TeV}, \ref{tab:corrwolf2TeV} and \ref{tab:corrwolf5TeV}, respectively.

As mentioned earlier, in the presence of new contributions the CKM element $V_{ij}$ is modified to  $V_{ij}^{\prime}= V_{ij}(1+\Delta_{NP})$. To check the impact of the NP on the extracted values of the CKM elements, we have extracted $V_{ij}$ in the fit with $\Delta_{NP}=0$ and compared them with the extracted values obtained from the fit results with $\Delta_{NP}\ne 0$. The numerical estimates for all nine CKM parameters in all the fit scenarios are given in Table.~\ref{tab:CKM}. Each of the numbers corresponds to the median and $1\sigma$ quantiles for the respective distributions of the CKM parameters. As expected, the extracted values remain unaltered in the presence of the NP effects we are considering.

As discussed in sub-section \ref{subsec:BtoDst}, we have analyzed the $B\to D^*\ell\nu_{\ell}$ ($\ell = e$ and $\mu$) decay mode independently and along with all the other inputs used to extract the Wolfenstein parameters. With the updated inputs from lattice, we carry out fits in the SM (without any new contribution) and include new contributions. In the frequentist and Bayesian analyses, the fit results for the BGL coefficients with and without $C_T$ are given in Table.~\ref{tab:bglfit}. For the semileptonic $P\to M$ decay modes we can define observables like $R(M^{(*)}) = \frac{\mathcal{B}(P \to M^{(*)}\tau\nu_{\tau})}{\mathcal{B}(P \to M^{(*)}\ell\nu_{\ell})}$. In the SM, these observables are expected to respect lepton-universality (LU), which can be violated (LUV) in the presence of new interactions affecting these decays. For the type of new effects we are considering here, the NP effects will cancel along with the CKM elements in $R(M)$.  However in $R(M^*)$, the new contributions will be affecting the decay rate distributions along with the vertex factor and the contribution will be sensitive to the lepton mass. Therefore, for $R(D^*)$, the new effects will not get cancelled completely. We also take this opportunity to update the SM prediction for $R(D^*)$ with the newly available inputs. Using the results given in Table.~\ref{tab:bglfit} along with the respective correlations, we have predicted $R(D^*)$ in the SM and in NP scenarios with three masses which are shown in Table.~\ref{tab:RDst}. The SM predictions are unchanged due to NP in $B\to D^*\ell\nu_{\ell}$ which are tightly constrained from the CKM fit analysis.

As we have mentioned earlier, the extraction of $C_T$ from the detailed analysis of the inclusive $B\to X_c\ell\nu_{\ell}$ decays is beyond the scope of this paper. Inspite of the difficulties in the extraction of new physics parameters from inclusive measurements, we have attempted to naively extract the allowed range of $C_T$ from the respective decay rate. Following the simplified approach as discussed in Ref.~\cite{Jung:2018lfu}, we define the approximate inclusive branching fraction in the presence of leading order new physics effect only in the rates

\beq
\mathcal{B}(B \to X_c e \nu)_{exp} \approx \mathcal{B}(B \to X_c e \nu)_{SM} \frac{\Gamma(B \to X_c e \nu)^{LO}_{NP}}{\Gamma(B \to X_c e \nu)^{LO}_{SM}}
\label{eq:NP-incl}
\eeq
where $\mathcal{B}(B \to X_c e \nu)_{exp} = (10.8 \pm 0.4)\%$ \cite{PDG2020} is the experimentally meaured branching fraction. The expressions for $\Gamma(B \to X_c e \nu)^{LO}_{NP}$ and $\Gamma(B \to X_c e \nu)^{LO}_{SM}$ can be found in \cite{Jung:2018lfu}. Note that $\Gamma(B \to X_c e \nu)^{LO}_{NP}$ is sensitive to $C_T$. We have expressed $\mathcal{B}(B \to X_c e \nu)_{SM}$ as given below 
\beq \label{eq:inclvcb}
\mathcal{B}(B \to X_c e \nu)_{SM} = \tau_B |V_{cb}|_{incl}^2\ \Gamma^\prime .
\eeq
Here, $\tau_B$ is the mean life of the B-meson, and $\Gamma^{\prime}$ is the integrated rate and includes all the available higher-order perturbative corrections up to $\mathcal{O}(\alpha_s^3)$ and the corrections to the non-perturbative matrix elements; for details see \cite{Alberti:2014yda,Bordone:2021oof}. Note that $\Gamma^{\prime}$ is a function of $m_b^2$, $m_c^2$, and different non-perturbative matrix elements which could be extracted from a fit to the respective leptonic energy moments and the moments of the hadronic invariant mass of the respective differential distributions \cite{Gambino:2013rza,Alberti:2014yda}. Using these fitted parameters, one can find out $\Gamma^{\prime}$ and thereby extract $|V_{cb}|$ from eq.~\ref{eq:inclvcb} by letting $\mathcal{B}(B \to X_c e \nu)_{SM} = \mathcal{B}(B \to X_c e \nu)_{exp}$. Following this approach the authors of ref.~\cite{Bordone:2021oof} have obtained $\Gamma^\prime = 2.44(11)\times 10^{-11}$ GeV, and hence $|V_{cb}|_{incl} = 42.16(51)\times 10^{-3}$. Note that the fit in ref.~\cite{Bordone:2021oof} does not assume contributions from NP.

\begin{figure}[t]
	\centering
	\includegraphics[scale=0.8]{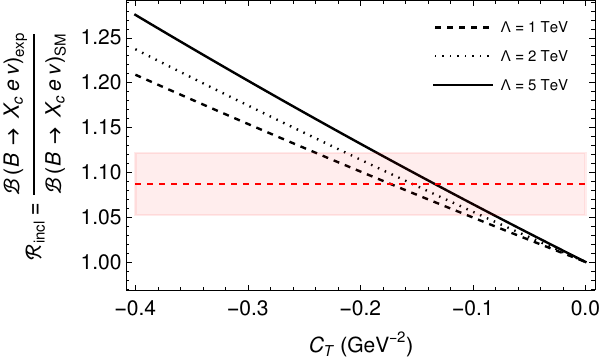}
	\caption{(Left) Variation of $\mathcal{R}_{incl}$ as a function of new physics coupling $C_T$ for three different values of scale $\Lambda$. The red dashed line represents the central value of the ratio with the $3\sigma$ uncertainty depicted by the shaded region. (Right) Variation of $|C_T|$ with $\Lambda$ within the $3\sigma$ allowed range of $\mathcal{R}_{incl}$.}
	\label{fig:CT-incl}
\end{figure}

In this analysis, we aim to constrain the magnitude of $C_T$ from eq.~\ref{eq:NP-incl} with the measured value $\mathcal{B}(B \to X_c e \nu)_{exp}$, and the $|V_{cb}|$ obtained from the CKM fit without any inputs from the inclusive decays. Ideally, one should simultaneously fit the $|V_{cb}|$ along with $C_T$; however, for the naive estimate, the only available input is the inclusive branching fraction. Therefore, we must fix $|V_{cb}|$ from the other measurements.
In order to extract the allowed range of $C_T$, we calculate the ratio 
\beq
\mathcal{R}_{incl} = \frac{\mathcal{B}(B \to X_c e \nu)_{exp}}{\mathcal{B}(B \to X_c e \nu)_{SM}} = 1.087 \pm 0.034.
\label{eq:Rincl}
\eeq
With the fitted $\Gamma^\prime = 2.44(11)\times 10^{-11}$ GeV, along with our predicted $|V_{cb}| = 0.04045^{+0.00038}_{-0.00037}$ from the CKM fit (Table.~\ref{tab:CKM}), we find $\mathcal{B}(B \to X_c e \nu)_{SM} = (9.9\pm 0.5)\%$ which is consistent with the measured value $\mathcal{B}(B \to X_c e \nu)_{exp}$ at 68\% C.L. Note that from eq.~\eqref{eq:NP-incl} we can define $\mathcal{R}_{incl}  \approx \frac{\mathcal{B}(B \to X_c e \nu)^{LO}_{NP}}{\mathcal{B}(B \to X_c e \nu)^{LO}_{SM}}$ and using the estimate in eq.~\ref{eq:Rincl} we can find out the allowed ranges of $C_T$. In Fig.~\ref{fig:CT-incl}, we have shown the variation of $\mathcal{R}_{incl}$ with $C_T$, from where we can get the allowed range of $C_T$ required to explain $\mathcal{R}_{incl}$ in its estimated range. The figure shows the estimated $1\sigma$ limit of the ratio $\mathcal{R}_{incl}$, and the allowed magnitude of $C_T$ could be as large as 0.2. As expected, the inclusive measurement does not provide a strong constraint on $C_T$, and a zero-consistent range is obtained even though large negative values of $C_T$ are favoured.

\subsection{DM phenomenology}

\begin{figure}[t]
	\centering
	\includegraphics[scale=0.82]{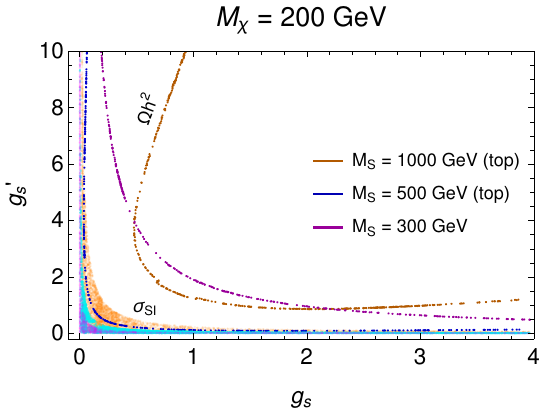}
	\includegraphics[scale=0.75]{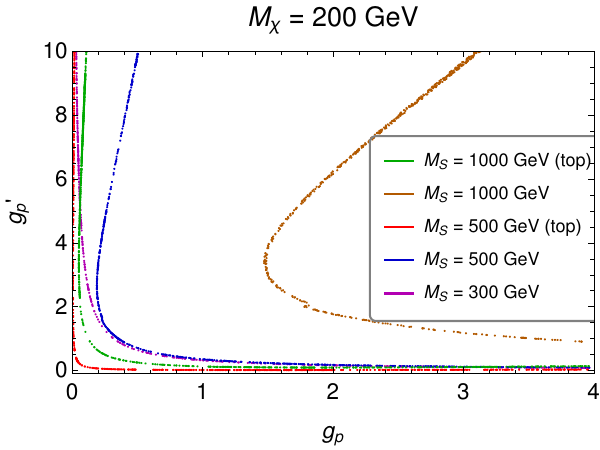}
	\caption{We show the parameter space in $g_s - g_s^\prime$ plane satisfying the DM relic constraints for mediator of masses $0.3$ TeV (dark magenta), $0.5$ TeV (dark blue), and $1$ TeV (dark orange) for $M_\chi = 200$ GeV in absence of pseudoscalar couplings on the left figure. The dark matter is under-abundant in the region on the right of these curves. We also show the region allowed by the Xenon-1T data on the spin-independent dark matter nucleon scattering cross-section in the lighter shades of the same colors. A similar plot in the $g_P-g_P^\prime$ plane in absence of scalar couplings is shown on the right.}
	\label{fig:gs-gsp}
\end{figure}

 We point out the main results from DM phenomenology in this section. Before we present our results related to DM phenomenology, we would like to point out different contributions to the relic density in the different regions of the masses of the DM ($M_{\chi}$), the mediator ($M_s$) and quarks, respectively. This exercise will be important given the mass hierarchy between $M_{\chi}$, $M_s$ and the top quark mass. In the simplified dark matter model we are considering, for $M_{\chi} < M_s < m_t $, the dominating annihilation channels which will contribute to relic are given by $\chi\bar{\chi} \to b\bar{b}, c\bar{c}, gg$, respectively. Here, the $\chi\bar{\chi} \to gg$ is a one-loop process where the dominant contribution to the effective $S \to gg$ vertex will be from a top quarks loop \cite{Arina:2016cqj,Haisch:2015ioa}. However, for $M_{\chi} > m_t$, the dominating annihilation channel is $\chi\bar{\chi} \to t\bar{t}$ due to quark mass dependence of mediator-quark couplings. At and above the top threshold, i.e. when $M_s \gtrsim 2 m_t$, resonant annihilation into the top-quark pairs is sufficient to generate the observed relic abundance. Finally, when $M_{\chi} > M_s$, the annihilation will dominantly proceeds via t-channel exchange $\chi\bar{\chi} \to SS$. Also, in a situation when $M_{\chi} > m_t, M_s$ and $M_s > 2 m_t$, there will be contribution to relic from both the s-channel $\chi\bar{\chi} \to t\bar{t}$ and the t-channel $\chi\bar{\chi} \to SS$ annihilation.
	
Considering the facts discussed above, in the context of DM phenomenology, we presented the analysis in two segments for a better understanding : one when $M_s < 2 m_t$ and the other for $M_s \gtrsim 2 m_t$.  We first begin by showing the allowed parameter space satisfying the relic data in presence of scalar couplings only i.e when $g_p = g_p^\prime = 0$ and $g_s, g_s^\prime \neq 0$ in the left plot of Fig~\ref{fig:gs-gsp}. The allowed values of $g_s$ can be inferred from the fit result of $C_T$ (with $C_p=0$) as tabulated in~\ref{tab:comb-NP-rslt}. For simplicity, we only consider the allowed solutions for $C_T$ that are positive in the $2\sigma$ range of the best fit estimate for the DM analysis. Therefore, we have $g_s = \sqrt{v^2 C_T}$ while $g_s^\prime$ is varied between $0-10$. Note the allowed value of $g_s$ is restricted to $\lsim 4$. In the left plot in Fig.~\ref{fig:gs-gsp}, we show the correlation between the two scalar couplings satisfying the relic abundance of dark matter of mass 200 GeV and mediator masses $300$, $500$ \& $1000$ GeV in dark magenta, dark blue and dark orange lines, respectively. However, we find that the scalar couplings required to satisfy the relic is ruled out from the spin-independent direct detection (SIDD) constraints from XENON-1T experiment \cite{XENON:2018voc}. The region allowed by the XENON data for the same DM and mediator masses as above are shown in lighter magenta, light blue and light orange respectively in the plot which are much away from the relic allowed curves. Hence, for all non-zero values of the scalar couplings $g_s, g_s^\prime$, the SIDD bound plays the most stringent role in constraining their upper limits. The same conclusion holds for values of DM masses other than $M_{\chi}=200$ GeV. Therefore, one can re-establish the fact that scalar portal DM candidates are not favoured by the data.

\begin{figure}[t]
	\centering
	\includegraphics[scale=0.8]{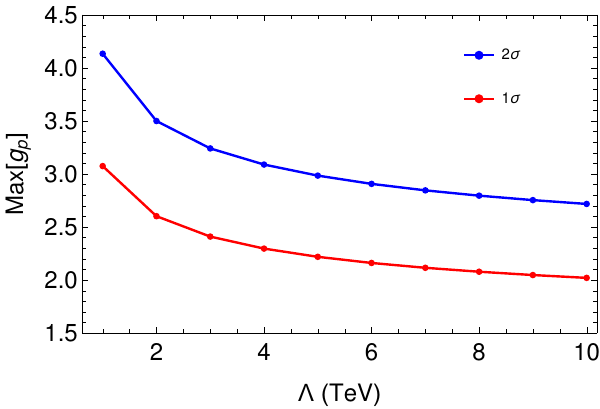}
	\caption{The variation of the maximum 1-$\sigma$ and 2-$\sigma$ allowed values of the coupling $g_p$ when $g_s \approx 0$ or negligibly small, with the cutoff scale.}
	\label{fig:gpvsLambda}
\end{figure}

Similarly, we show the correlations between the pseudoscalar couplings $g_p$ and $g_p^{\prime}$ on the right side of Fig.~\ref{fig:gs-gsp} for similar DM and mediator masses. Here, we have set $g_s = g_s^{\prime} =0$ and obtained the bound on $g_p$ from the results of table \ref{tab:comb-NP-rslt} using the relation $g_p = \sqrt{v^2 C_T}$ while $g_p^\prime$ is varied between $0-10$. In Fig.~\ref{fig:gpvsLambda}, we have shown the variation of the maximum of the 1-$\sigma$ and 2-$\sigma$ allowed values of the coupling $g_p$ (with $g_s \approx 0$) with the cut-off scale $\Lambda$. There are slight reductions in the allowed upper limit of the coupling with the increasing values of the cutoff scale for $\Lambda \lsim 5$ TeV. The changes in the allowed values of the coupling are almost negligible for $\Lambda > 5$ TeV. This is due to the logarithmic dependence of the coupling on the cutoff scale. Note that for the mediator masses below the top threshold, as an example for $M_S = 300$ GeV the allowed values of $g_p^{\prime}$ is $\lsim 1$ and its value could be $> 1$ only if $g_p << 1$. Also, for $g_p \gsim 0.5$ the allowed values of $g_p^{\prime}$ will be $<< 1$. The constraints are severe for the values of $M_S$ larger than $2 m_t$ (top-threshold). Such an observation is also true for the scalar coupling case discussed in the previously. We have shown the correlation between the couplings for $M_S = 1000$ and $1500$ GeV. It should be noted that, if kinematically allowed, i.e. for $M_S > 2m_t$, the DM can annihilate into a pair of top quarks whose cross-sections are enhanced due to the heavy top mass effect in the interaction (see eq.~\ref{eq:Lag2}). If we neglect the top quark interaction, the resulting correlation is presented by the blue and brown curves on the right plot of Fig.~\ref{fig:gs-gsp} for $M_S = 0.5$ and 1 TeV respectively. Larger couplings become allowed if annihilation to only light quarks is considered.  
In the absence of the scalar couplings, the parameter space is free from any constraints from the DM-nucleon scattering data since the pseudoscalar couplings give rise to velocity suppressed spin-dependent scattering. 

\begin{figure}[t]
\centering
\includegraphics[scale=0.64]{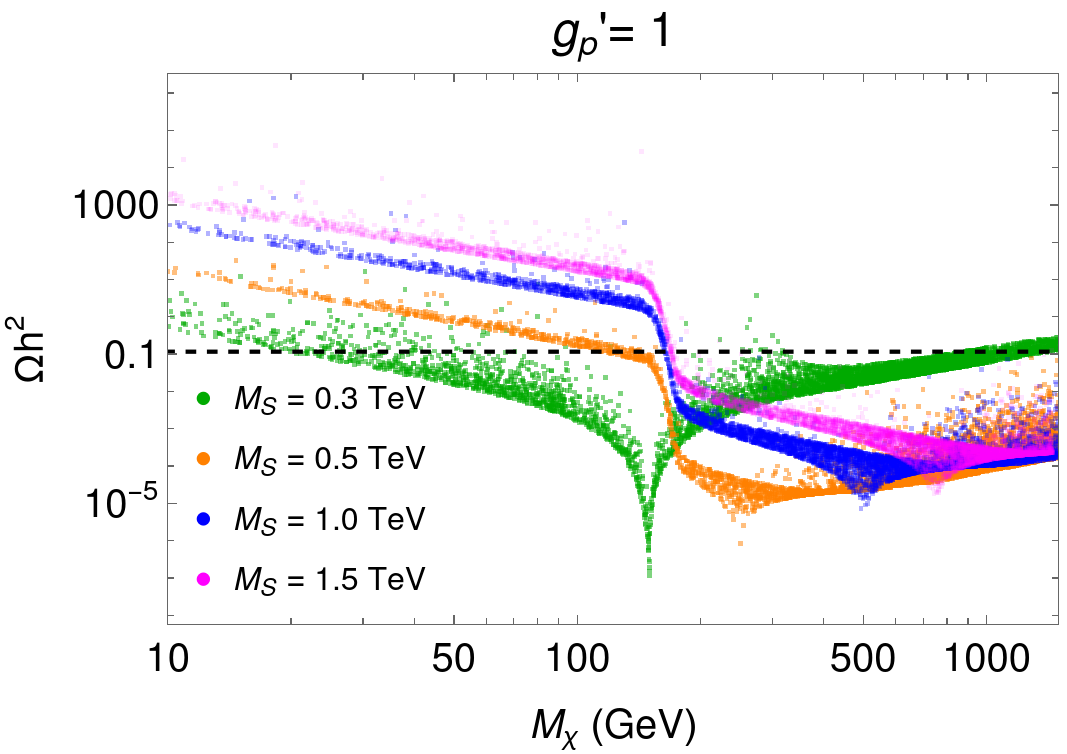}~~
\includegraphics[scale=0.65]{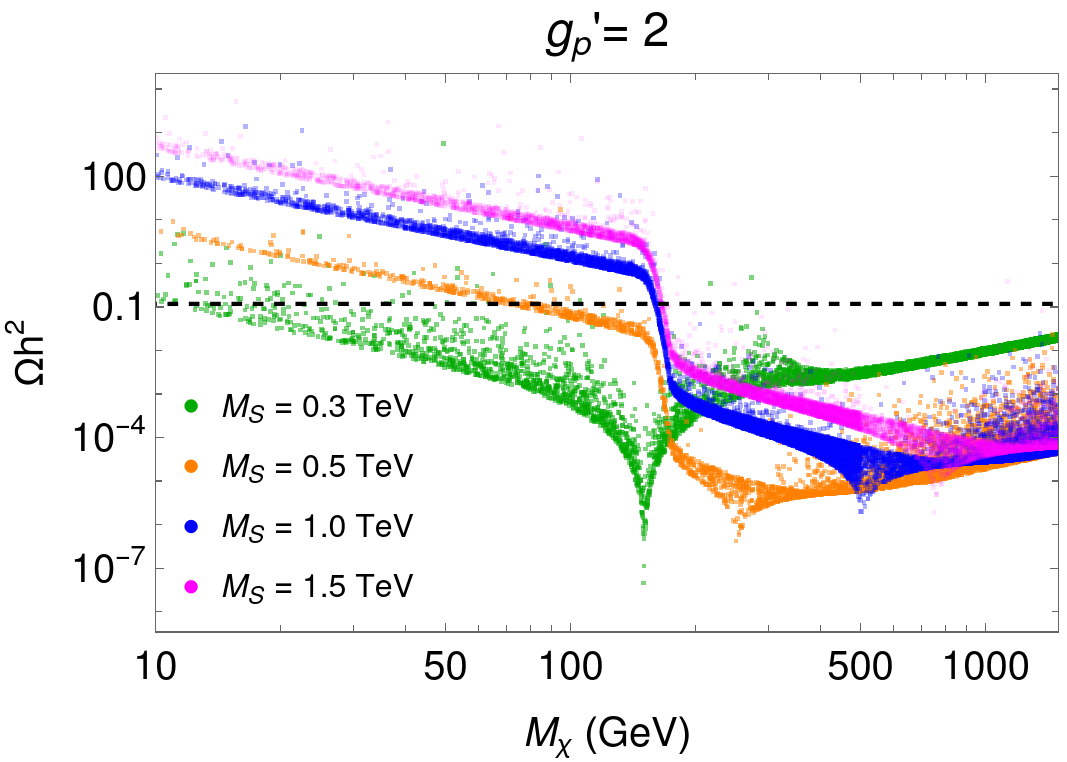}
\caption{Here we plot the relic abundance as a function of the DM mass for the four values of $M_S$, as denoted by the green, orange, blue and magenta points, for $0\leq g_P \leq 4$ and $g_{p^\prime}$ taken to be 1 (left plot) and 2 (right plot). See text for more details.}
\label{fig:relicVsMDM}
\end{figure}

\begin{figure}[t]
\centering
\includegraphics[scale=0.81]{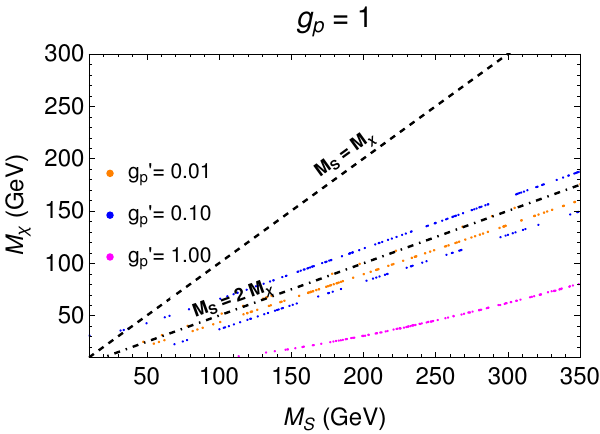}
\includegraphics[scale=0.83]{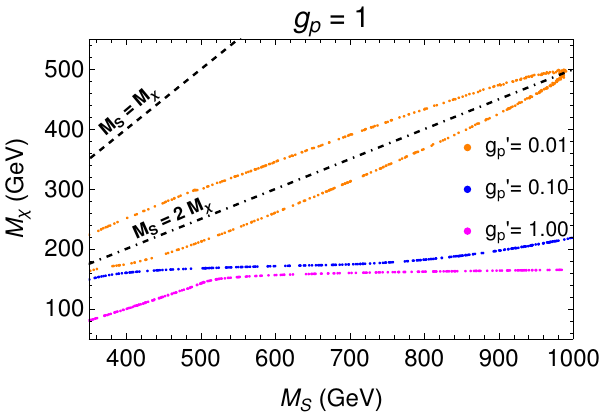}
\caption{The relic allowed regions in $M_\chi - M_S$ plane for different values of pseudoscalar couplings when $M_S < 2 m_t$ (left) and $M_S > 2 m_t$ (right).}
\label{fig:MDM-MS}
\end{figure}

So far, we have focused on a particular mass of the DM. We will discuss the phenomenology for the varying mediator and DM masses in what follows. The variation of the relic abundance with the DM mass is shown in Fig.~\ref{fig:relicVsMDM}. Here, we show the variation with the DM mass for $M_S = 0.3, 0.5 , 1.0$ and $1.5$ TeV, in green, orange, blue and magenta points respectively, for the scenario $ 0 \le g_p \le 4 $ (as obtained above) and $g_p^\prime = 1$ (left), 2 (right). The black dashed line signifies the present-day relic abundance of DM. For $g_p^\prime = 1$, from the scans, we find that for relatively lighter masses, like $M_S = 300$ GeV, the correct relic is satisfied for a wide range of DM masses on both sides of $M_S \approx 2 M_{\chi}$. However, for the values of $M_S$ higher than the top-threshold, the allowed DM masses become very constrained to a small region, and they satisfy $M_{\chi} << M_S/2$. A similar plot for $g_p^\prime = 2$ is shown in the right panel of Fig.~\ref{fig:relicVsMDM}. One can note that the probability of getting a allowed solution for $M_{\chi} > M_S/2$ has reduced as compared to that for $g_p^\prime = 1$. Therefore, for higher values of the coupling $g_p^{\prime} (> 2)$, the relic will be satisfied only when $M_{\chi} <  M_S/2$. On the other hand, one would expect the solution to be concentrated near the $M_{\chi} \approx  M_S/2$ for values of $M_S$ near or below the top threshold and $g_p^\prime < 1$.

In Fig.~\ref{fig:MDM-MS}, we show the allowed correlations between the mediator and the dark matter mass for a few values of $g_{p^\prime}$ while we fix $g_p = 1$. The relic abundance is satisfied only along the orange, blue and magenta curves for $g_{p^\prime} = 0.01, 0.1$ and $1.0$ respectively. These benchmark values are chosen after an inspection of the correlations in Fig.~\ref{fig:gs-gsp}. The plot on the left is for the mediator mass $M_S \lsim 2 m_t$ GeV, and that to the right is for above it. The black dashed and dot-dashed lines depict $M_\chi = M_S$ and $M_\chi = 2 M_S$ respectively. It is evident from the plots that for $M_S < 2 m_t$, for lower values of $g_p^\prime (\lsim 1)$, relic is satisfied near the region $M_{\chi} \approx  M_S/2$. However, for a relatively high value, like $g_p^\prime \approx 1$ the relic will be satisfied only when $M_{\chi} <  M_S/2$. For $M_S > 2 m_t$, for smaller values of $g_p^{\prime}$, as an example for $g_p^{\prime} = 0.01$, the relic will be satisfied near the region $M_{\chi} \approx M_S/2$. However, as observed earlier, the relic will be satisfied for $M_{\chi} <  M_S/2$ when $g_p^{\prime}$ is not too small. High DM masses ($\gsim 350$ GeV) do not satisfy the relic constraint until the annihilation to the top final state becomes relevant. This is expected since the annihilation cross-section is lowered for low values of the couplings, thereby making the relic overabundant. Therefore, we need smaller DM masses to tune the relic to the correct amount by increasing the overall cross-section. The reverse is true for the cases shown on the right plot of Fig.~\ref{fig:MDM-MS} since the top quark effect is so dominant that high DM masses are disfavoured unless the couplings are very low.  

Based on the analyses of 8 TeV data, the relevant bounds on $g_s, g_s^{\prime}, g_p, g_p^{\prime}, M_S$ and $M_{\chi}$ from the collider searches: $\slashed{E_T} + jet$ and $\slashed{E_T} + \bar{t}t$, respectively, can be seen from the refs.~\cite{Arina:2016cqj, Haisch:2015ioa}. The corresponding bounds have been updated in the newly available analyses from ATLAS \cite{ATLAS:2021kxv} and CMS \cite{CMS:2018prl,CMS:2018rkg,CMS:2019long,CMS:2021far}, which are based on the data at the centre of mass energy $\sqrt{s} = 13$ TeV. Though a dedicated collider analysis is beyond our paper's scope or motivation, we can draw some inferences from the results of ATLAS and CMS on the respective model parameters. The ATLAS and CMS have looked for signatures in $X + \slashed{E}_T$ final states where $X$ stands for $W/Z/\gamma$ or jets. Note that the LHC bounds are applicable only for $M_s \gtrsim 2 M_{\chi}$. As was seen in these experimental analyses, the strong constraints on the scalar or pseudoscalar mediator masses come mainly from the analyses of $(t/\bar{t}/t\bar{t} + \slashed{E}_T)$ \cite{CMS:2018prl,CMS:2019long} and $\slashed{E_T} + jets$ signatures \cite{ATLAS:2021kxv,CMS:2021far}. As can be seen from ref.~\cite{CMS:2019long}, from the analysis of combined events in $(t/\bar{t}/t\bar{t} + \slashed{E}_T)$ searches for $g_{p} = g_{p}^{\prime} = 1$ and the DM mass $M_{\chi} = 1$ GeV, mediator masses $M_s < 300$ GeV are excluded at 95\% CL. On the other hand, at both CMS and ATLAS \cite{ATLAS:2021kxv,CMS:2021far}, from the searches of the energetic jets plus large missing transverse energy (MET), constraints are obtained on $M_s$ for $M_{\chi} = 1$ GeV and with the magnitude of the mediator couplings: $g_{p} \approx 1.5$ and $ g_{p}^{\prime} = 1$ \footnote{Note that in the refs.~\cite{ATLAS:2021kxv,CMS:2021far} the couplings of the mediator with the quarks have been defined a little differently than we defined it in our paper. The couplings are related by the relation $g_p = \frac{v}{m_t} g_{LHC} $ where $m_t$ is the mass of the top quark. However, in ref.~\cite{CMS:2019long}, the definition of $g_{p}$ or $g_s$ exactly matches ours.}. Their studies exclude the pseudoscalar mediator mass $ M_s$ up to 470 GeV at 95\% CL. As proposed in ref.~\cite{LHCDMgroup:2019report}, the constraints are obtained in the $M_s - M_{\chi}$ plane from the LHC analyses for the fixed values of the mediator couplings, which can be seen from \cite{ATLAS:2021kxv}, and it could be helpful to exclude a part of the region of low DM masses.   	

In addition, we understand that the jets plus MET production cross section will increase with the increasing values of $g_p$ or $g_p^{\prime}$. One should note that the maximum excluded values of $M_s$ decrease with increasing $M_{\chi}$, as the branching fraction of the mediator to DM particle decays diminishes \cite{LHCDMgroup:2019report}. Therefore, with increasing values of the DM masses, the bounds on the mediator masses will be less stringent. Also, the naive expectation is that the mediator production cross sections will decrease with the increasing values of $M_s$. Therefore, it is expected that for higher values of the $M_{\chi}$ and $M_s$ ($\gtrsim 470$ GeV), the mediator coupling $g_p > 1.5$ will be allowed by the LHC data. Also, the current LHC bound on $M_s$ will be relaxed for $g_{p}$ or $g_{p}^{\prime} < 1$. In our analysis, we obtained the relic allowed regions for $g_{p} = g_{p}^{\prime} = 1$ shown in colour magenta in Fig.~\ref{fig:MDM-MS} for $M_s \lesssim 350$ GeV (left plot) which will not be allowed by the LHC data. In the same figure, the other allowed solutions are shown for $g_{p}^{\prime}= 0.1$ and $0.01$, respectively, which might still be allowed by the current experimental searches. For $g_p^{\prime}=0.1$ or $0.01$, due to the reduction in production cross-section, the exclusion limit on $M_s$ will be less stringent than what has been obtained for $g_p^{\prime}=1$. Also, the allowed regions shown in the right plot of Fig.~\ref{fig:MDM-MS} for $M_s > 400$ GeV will be allowed by the current LHC limits. A dedicated collider analysis may be required to support these arguments, which is beyond the scope of this paper. Also, the current collider searches have not ruled out the upper limits on the values of $g_s$ or $g_p$ obtained from the CKM fit. The variations of the maximum allowed value $g_p$ with the cut-off scale have been shown in Fig.~\ref{fig:gpvsLambda}, a more stringent bound from the future collider experiment may be helpful to get some understanding about the viable cut-off scale. More precise data might be helpful to constrain it further both at the collider and in the CKM fit.

\section{Summary}
From the global CKM fit analysis, this paper analyzes the constraints on the parameters of a class of NP models having neutral quark current interaction mediated by a heavy scalar. This kind of NP has an impact on the leptonic and semileptonic decays at the one-loop level. Also, with the newly available updates, we have extracted Wolfenstein parameters and the related CKM elements with and without a contribution from NP from the global fit. In this paper, we mainly focus on the impact of our bounds on DM phenomenology. However, the bounds might be applicable in any other relevant phenomenology.    

We have considered a simplified DM model with fermionic dark matter whose interactions with the SM is mediated by a heavy neutral scalar. There is no symmetry to forbid the interactions of the SM quarks to this new scalar. Hence, it will contribute to the charged current vertices of $\bar{d}_i u_j W$ at one loop level. The modifications to the $P \to M$ and $P \to M^*$ transitions due to the new interactions are quite contrasting. In case of the leptonic $P \to \ell \nu_\ell$ and semileptonic $P \to M$ decays, the vertex factors will be altered while in case of the $P \to M^*$ semileptonic decays, the $q^2$ decay distribution itself is modified. As a recent developement, lattice results on the form factors of the $B \to D^* \ell \nu_\ell$ decay at non-zero recoil are now available. Therefore we update the SM prediction of the CKM element $|V_{cb}|$ before incorporating the NP effects. We obtain $|V_{cb}| = 38.69(79) \times 10^{-3}$ at $68\%$ CL. We also predict the observable $R(D^*)$ in the different fit scenarios with and without the NP.

With this new update and all other available CKM measurements, we perform a global fit in the presence of the NP effects for some fixed values of the mediator mass. From this fit, we can only constrain the combination $C_T$ and not the individual couplings $C_s, C_p$. We show that for high values of $M_S$, the coupling gets severely constrained from the data. From the dark matter SIDD constraints, we can restrict the scalar couplings $C_s$ and $g_s^\prime$ to minimal values. This automatically translates to a bound on the parameter $C_p$ from our fit results on $C_T$. However, since the pseudoscalar couplings have velocity suppressed contribution to the spin-dependent DD cross-section, there remains some freedom in $g_p^\prime$. With this setup, we have discussed the relevant DM phenomenology. 

\acknowledgments

This work of SN is supported by the Science and Engineering Research Board, Govt. of India, under the grant CRG/2018/001260. 

\appendix

\section{Fit results for the BGL coefficients }\label{BGL}
\begin{figure}[ht]
	\centering
	{\includegraphics[scale=0.36]{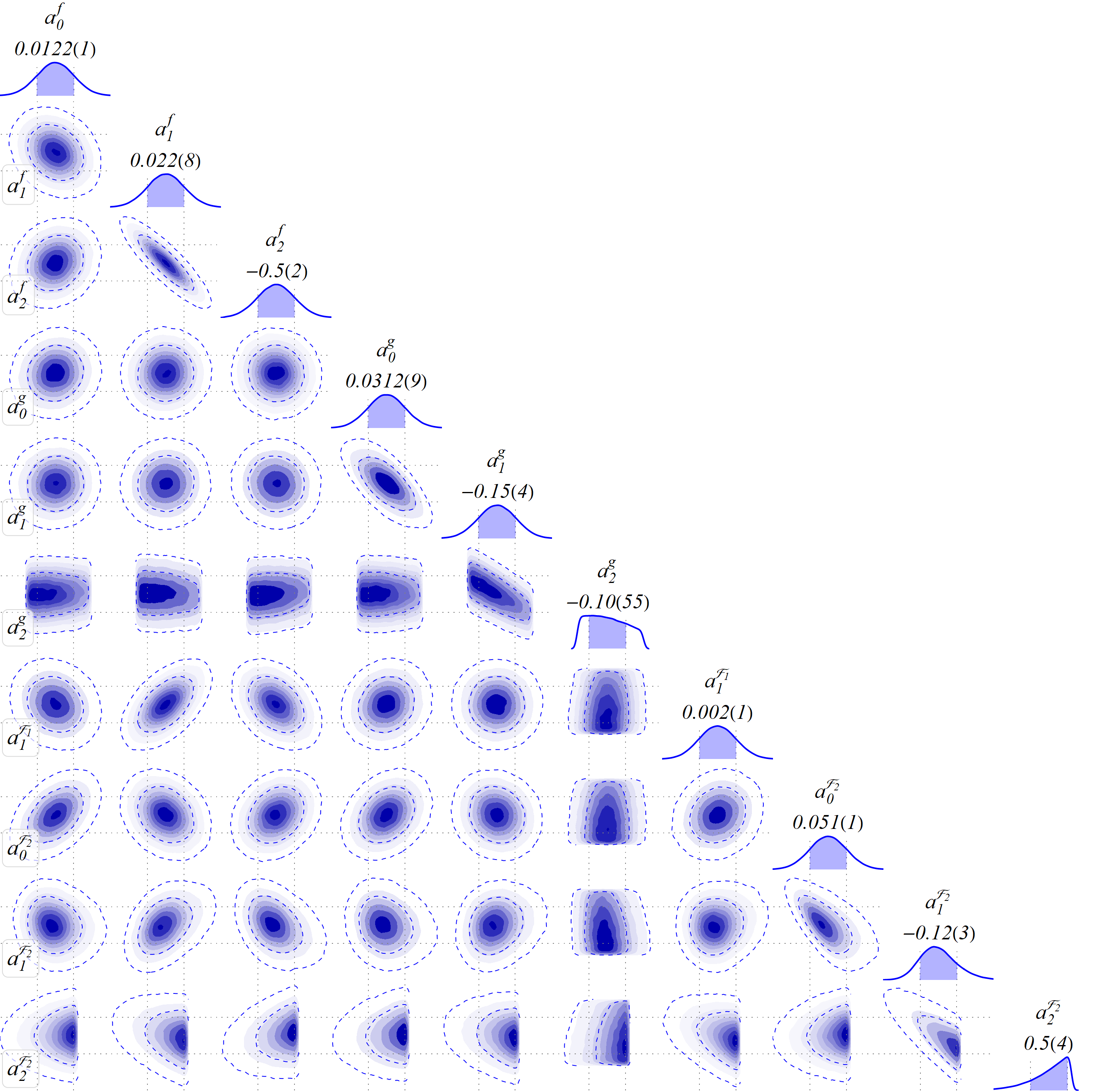}}
	\caption{The triplot for the BGL parameters for the SM. We checked and found out that there are no appreciable changes for the three NP scenarios as compared to the SM as far as the posterior and correlations for the BGL parameters are concerned. The central value and corresponding errors for the parameters are provided at the top of the corresponding 1-D posteriors.}
	\label{fig:triplot_BGL}
\end{figure}

The fitted values of the BGL coefficients (table \ref{tab:bglfit} ) defined in eq. \ref{eq:FF-BGL} which are obtained from a combined fit to $B\to D^*\ell\nu_{\ell}$ decay rates and other relevant inputs used in global CKM fit analysis. In Fig. \ref{fig:triplot_BGL}, we provide the triplot for the BGL parameters corresponding to the SM in this section which are almost unchanged in the presence of NP.

	\begin{sidewaystable}[ht]
		\centering
		\small
		\begin{tabular}{|c|c|c|c|c|c|c|c|c|}
			\hline
			\multirow{2}{*}{\textbf{Parameters}} & \multicolumn{4}{c|}{\textbf{Frequentist}} & \multicolumn{4}{c|}{\textbf{Bayesian}}\\
			\cline{2-9}
	   &  $\text{SM}$  &  $\text{$\Lambda $=1 TeV}$  &  $\text{$\Lambda $=2 TeV}$  &  $\text{$\Lambda $=5 TeV}$ &  $\text{SM}$  &  $\text{$\Lambda $=1 TeV}$  &  $\text{$\Lambda $=2 TeV}$  &  $\text{$\Lambda $=5 TeV}$\\\hline
	$a_0^f$  &  $\text{0.01219$\pm $0.00012}$  &  $\text{0.01219$\pm $0.00012}$  &  $\text{0.01219$\pm $0.00012}$  &  $\text{0.01219$\pm $0.00012}$  &  $\text{0.01218$\pm $0.00012}$  &  $\text{0.01218$\pm $0.00012}$  &  $\text{0.01218$\pm $0.00012}$  &  $\text{0.01219$\pm $0.00012}$  \\
  $a_1^f$  &  $\text{0.0203$\pm $0 .0092}$  &  $\text{0.0202$\pm $0.0092}$  &  $\text{0.0202$\pm $0 .0092}$  &  $\text{0.0202$\pm $0.0092}$  &  $\text{0.0222$\pm $0.008}$  &  $0.0221_{-0.0081}^{+0.008}$  &  $0.022_{-0.008}^{+0.0081}$  &  $0.0214_{-0.0079}^{+0.0077}$  \\
  $a_2^f$  &  $\text{-0.49$\pm $0.19}$  &  $\text{-0.49$\pm $0.19}$  &  $\text{-0.49$\pm $0.19}$  &  $\text{-0.49$\pm $0.19}$ &  $\text{-0.53$\pm $0.17}$  &  $\text{-0.53$\pm $0.17}$  &  $\text{-0.52$\pm $0.17}$  &  $\text{-0.51$\pm $0.16}$  \\
  $a_0^g$  &  $\text{0.0313$\pm $0.00095}$  &  $\text{0.0313$\pm $0.00095}$  &  $\text{0.0313$\pm $0.00095}$  &  $\text{0.0313$\pm $0.00095}$  &  $\text{0.03121$\pm $0.00094}$  &  $\text{0.03122$\pm $0 .00094}$  &  $\text{0.03121$\pm $0.00094}$  &  $0.03124_{-0.00094}^{+0.00095}$  \\
  $a_1^g$  &  $\text{-0.142$\pm $0.062}$  &  $\text{-0.143$\pm $0.062}$  &  $\text{-0.143$\pm $0.062}$  &  $\text{-0.143$\pm $0.062}$  &  $\text{-0.149$\pm $0.036}$  &  $-0.147_{-0.037}^{+0.035}$  &  $-0.146_{-0.037}^{+0.035}$  &  $-0.154_{-0.036}^{+0.037}$  \\
  $a_2^g$  &  $\text{-0.43$\pm $1 .44}$  &  $\text{-0.41$\pm $1.44}$  &  $\text{-0.41$\pm $1.44}$  &  $\text{-0.41$\pm $1.44}$ &  $-0.13_{-0.58}^{+0.68}$  &  $-0.22_{-0.54}^{+0.7}$  &  $-0.25_{-0.49}^{+0.77}$  &  $0.022_{-0.685}^{+0.627}$  \\
  $a_1^{\mathcal{F}_1}$  &  $\text{0.0017$\pm $0.0014}$  &  $\text{0.0017$\pm $0.0014}$  &  $\text{0.0017$\pm $0.0014}$  &  $\text{0.0017$\pm $0.0014}$  &  $\text{0.0022$\pm $0.0012}$  &  $\text{0.0022$\pm $0.0012}$  &  $\text{0.0022$\pm $0 .0012}$  &  $\text{0.0021$\pm $0.0012}$  \\
  $a_ 0^{\mathcal{F}_2}$  &  $\text{0.0508$\pm $0.0012}$  &  $\text{0.0508$\pm $0.0012}$  &  $\text{0.0508$\pm $0.0012}$  &  $\text{0.0508$\pm $0.0012}$  &  $0.0507_{-0.0011}^{+0.0012}$  &  $0.0507_ {-0.0011}^{+0.0012}$  &  $\text{0.0507$\pm $0.0012}$  &  $0.0508_{-0.0011}^{+0.0012}$  \\
  $a_1^{\mathcal{F}_2}$  &  $\text{-0.149$\pm $0.058}$  &  $\text{-0.149$\pm $0.058}$  &  $\text{-0.149$\pm $0 .058}$  &  $\text{-0.149$\pm $0.058}$  &  $-0.125_{-0.028}^{+0.033}$  &  $-0.127_{-0.028}^{+0.032}$  &  $-0.126_{-0.028}^{+0.033}$  &  $-0.134_{-0.026}^{+0.028}$  \\
  $a_2^{\mathcal{F}_2}$  &  $\text{0.99$\pm $0.9}$  &  $\text{0.99$\pm $0.9}$  &  $\text{0.99$\pm $0.9}$  &  $\text{0.99$\pm$0.9}$  &  $0.61_{-0.49}^{+0.28}$  &  $0.63_{-0.44}^{+0.27}$ &  $0.62_{-0.45}^{+0.28}$  &  $0.76_{-0.38}^{+0.18}$  \\\hline
		\end{tabular}
		\caption{SM and NP estimates for the BGL parameters. The NP estimates have been presented for both the cases with the NP scale $\Lambda=$ 1, 2 and 5 TeV. The Bayesian estimates correspond to the median and 1$\sigma$ Quantiles of the posteiors for the respective parameters.}
		\label{tab:bglfit}
	\end{sidewaystable}

\section{Correlations between the Wolfenstein parameters and $C_T$}\label{correlation}
In this section we provide numerical estimates for the correlations between $A$, $\lambda$, $\bar{\eta}$, $\bar{\rho}$ and $C_T$ corresponding to the analyses with and without any NP contributions. In case of NP, we have presented the results for the corresponding scale $\Lambda =$ 1 TeV and 2 TeV respectively. These have been obtained from the Bayesian posteriors.

\begin{table}[htp!!]
\begin{tabular}{|c|c|c|c|c|c|}\hline
	$\text{SM}$  &  $\text{A}$  &  $\lambda$  &  $\bar{\rho }$  &  $\bar{\eta }$  \\\hline
	$\text{A}$  &  $1.$  &  $-0.258465$  &  $-0.210554$  &  $-0.493578$  \\
	$\lambda$  &  $-0.258465$  &  $1.$  &  $0.0728912$  &  $-0.055793$  \\
	$\bar{\rho }$  &  $-0.210554$  &  $0.0728912$  &  $1.$  &  $0.409138$  \\
	$\bar{\eta }$  &  $-0.493578$  &  $-0.055793$  &  $0.409138$  &  $1.$  \\\hline
\end{tabular}
	\caption{Correlations between the four Wolfenstein parameters for corresponding to the fit without NP.}
	\label{tab:corrwolfSM}
\end{table}

\begin{table}[htp!!]
	\begin{tabular}{|c|c|c|c|c|c|}\hline
	$\text{$\Lambda $=1 TeV}$  &  $C_T$  &  $\text{A}$  &  $\lambda$  &  $\bar{\rho }$  &  $\bar{\eta }$  \\\hline
	$C_T$  &  $1.$  &  $-0.00380722$  &  $0.00172136$  &  $0.0017101$  &  $0.00245208$  \\
	$\text{A}$  &  $-0.00380722$  &  $1.$  &  $-0.260845$  &  $-0.216509$  &  $-0.492561$  \\
	$\lambda$  &  $0.00172136$  &  $-0.260845$  &  $1.$  &  $0.075102$  &  $-0.0563302$  \\
	$\bar{\rho }$  &  $0.0017101$  &  $-0.216509$  &  $0.075102$  &  $1.$  &  $0.405289$  \\
	$\bar{\eta }$  &  $0.00245208$  &  $-0.492561$  &  $-0.0563302$  &  $0.405289$  &  $1.$  \\\hline
	\end{tabular}
	\caption{Correlations between the four Wolfenstein parameters along with $C_T$ for NP scale $\Lambda=1$ TeV.}
	\label{tab:corrwolf1TeV}
\end{table}

\begin{table}[htp!!]
	\begin{tabular}{|c|c|c|c|c|c|}\hline
	$\text{$\Lambda $=2 TeV}$  &  $C_T$  &  $\text{A}$  &  $\lambda$  &  $\bar{\rho }$  &  $\bar{\eta }$  \\\hline
	$C_T$  &  $1.$  &  $-0.00117545$  &  $0.000992878$  &  $0.000133051$  &  $0.000899192$  \\
	$\text{A}$  &  $-0.00117545$  &  $1.$  &  $-0.260192$  &  $-0.214117$  &  $-0.492652$  \\
	$\lambda$  &  $0.000992878$  &  $-0.260192$  &  $1.$  &  $0.0719362$  &  $-0.0561466$  \\
	$\bar{\rho }$  &  $0.000133051$  &  $-0.214117$  &  $0.0719362$  &  $1.$  &  $0.414471$  \\\hline
	\end{tabular}
	\caption{Correlations between the four Wolfenstein parameters along with $C_T$ for NP scale $\Lambda=2$ TeV.}
	\label{tab:corrwolf2TeV}
\end{table}

\begin{table}[htp!!]
	\begin{tabular}{|c|c|c|c|c|c|}\hline
	$\text{$\Lambda $=5 TeV}$  &  $C_T$  &  $\text{A}$  &  $\lambda$  &  $\bar{\rho }$  &  $\bar{\eta }$  \\\hline
	$C_T$  &  $1.$  &  $-0.00275994$  &  $-0.000450748$  &  $0.00303057$ &  $0.00405182$  \\
  $\text{A}$  &  $-0.00275994$  &  $1.$  &  $-0.258676$ &  $-0.212262$  &  $-0.494841$  \\
  $\lambda$  &  $-0.000450748$  &  $-0.258676$  &  $1.$  &  $0.0715675$  &  $-0.0577073$  \\
  $\bar{\rho }$  &  $0.00303057$  &  $-0.212262$  &  $0 .0715675$ &  $1.$  &  $0.406381$  \\
  $\bar{\eta }$  &  $0.00405182$  &  $-0.494841$  &  $-0.0577073$ &  $0.406381$  &  $1.$  \\
\hline
	\end{tabular}
	\caption{Correlations between the four Wolfenstein parameters along with $C_T$ for NP scale $\Lambda=5$ TeV.}
	\label{tab:corrwolf5TeV}
\end{table}

\newpage
\section{Fit Results Including Belle 2017 Data}
\label{sec:Belle2017}
In this section we point out the effect on the fit results if we include the $B \to D^* \ell \nu$ data from Belle 2017 analysis \cite{Abdesselam:2017kjf}. In Table.~\ref{tab:BDstlnu-SM-2}, we list the results for the SM fit to the data for extracting $|V_{cb}|$ and BGL form factor parameters. We can see from the table taht the fit values are consistent with those obtained without considering Belle 2017 dataset given in Table.~\ref{tab:BDstlnu-SM}. Similarly, for the full CKM fit including new physics, the fit values, as shown in Table.~\ref{tab:comb-NP-rslt-2}, for $C_T$ remain unchanged while the Wolfenstein parameters remain consistent within the $1\sigma$ errors of the previous fit result.

\begin{table}[t]
	\centering
	\renewcommand{\arraystretch}{1.5}
	\begin{tabular}{|c|c|c|c|c|}
		\hline
		\multirow{2}{*}{Dataset} & \multicolumn{2}{|c|}{Fit Quality} & \multirow{2}{*}{Parameter} & \multirow{2}{*}{Fit Result} \\
		\cline{2-3} 
		& $\chi^2/dof$ & p-Value & &\\
		\hline
		& \multirow{11}{*}{$101.66/85$} & \multirow{11}{*}{$10.5\%$} & $|V_{cb}|$ & $38.55 (70) \times 10^{-3} $ \\
		& & & $a_{0}^f$ & $0.0123 (1)$ \\
		& & & $a_{1}^f$ & $0.0192 (95)$ \\
		& & & $a_{2}^f$ & $-0.460 (194)$ \\
		\multirow{2}{*}{Belle'17  \cite{Abdesselam:2017kjf} + Belle'18 \cite{Belle:2018ezy} +} & & & $a_{0}^g$ & $0.0319 (10)$ \\
		$h_{A_1} (1)$ \cite{FermilabLattice:2014ysv}+ LCSR \cite{Gubernari:2018wyi} + Lattice \cite{FermilabLattice:2021cdg} & & & $a_{1}^g$ & $-0.140 (62)$ \\
		& & & $a_{2}^g$ & $-0.25 (145)$ \\
		& & & $a_{1}^{\mathcal{F}_1}$ & $0.0021 (15)$ \\
		& & & $a_{0}^{\mathcal{F}_2}$ & $0.0517 (12)$ \\
		& & & $a_{1}^{\mathcal{F}_2}$ & $-0.160 (59)$ \\
		& & & $a_{2}^{\mathcal{F}_2}$ & $0.986 (916)$ \\
		\hline
	\end{tabular}
	\caption{Similar to Table.~\ref{tab:BDstlnu-SM} including the dataset from Belle 2017 analysis  \cite{Abdesselam:2017kjf}.}
	\label{tab:BDstlnu-SM-2}
\end{table}

\begin{table}[h!!!]
	\renewcommand{\arraystretch}{1.5}
	\centering
	\resizebox{\textwidth}{!}{
		\begin{tabular}{|cc|c|c|c|c|c|c|c|}
			\hline
			\multicolumn{2}{|c|}{\multirow{2}{*}{$\text{Case}$}}  &  \multirow{2}{*}{$\chi^2\text{/dof}$}  &  \multirow{2}{*}{$\text{p-Value ($\%$)}$} &  \multicolumn{5}{c|}{$\text{Fit Result}$} \\
			\cline{5-9}
			&&  &  &  $C_T$ (GeV$^{-2}$)  &  $\text{A}$  &  $\lambda$ &  $\bar{\rho}$  &  $\bar{\eta}$  \\
			\hline
			\multirow{3}{*}{$\text{Scale} \Bigg \lbrace $ } & $1 \text{ TeV}$  &  $140.5/110$  &  $3$  &  $0.00003 \pm 0.00013$  & $0.795114 \pm 0.007472$  &  $0.224979 \pm 0.000293$  &  $0.17781 \pm 0.00973$  &  $0.390305 \pm 0.011842$  \\
			& $2 \text{ TeV}$  &  $140.5/110$  &  $3$  &  $0.00002 \pm 0.00009$  & $0.795112 \pm 0.007472$  &  $0.224979 \pm 0.000293$  &  $0.177811 \pm 0.009728$  &  $0.390305 \pm 0.011842$  \\
			& $5 \text{ TeV}$  &  $140.5/110$  &  $3$  &  $0.000015 \pm 0.000066$  & $0.795112 \pm 0.007472$  &  $0.224979 \pm 0.000293$  &  $0.177811 \pm 0.009728$  &  $0.390305 \pm 0.011842$  \\
			\hline
		\end{tabular}}
		\caption{Fit results similar to those in Table.~\ref{tab:comb-NP-rslt} by additionally including Belle 2017 dataset.}
		\label{tab:comb-NP-rslt-2}
	\end{table}

\newpage
\bibliography{CKMref} 
\end{document}